%% file: causal_perspective.tex
\documentclass[onefignum, onetabnum,onealgnum,onethmnum, oneeqnum]{siamonline190516}
\input{sf_preamble.tex}

\usepackage[symbol]{footmisc}

\usepackage{ericmath}
\usepackage[createShortEnv]{proofs}
\usepackage{xcolor}

\usepackage{tcolorbox}
\tcbuselibrary{breakable}
\usepackage{thmtools}
\usepackage{etoolbox}  

\usepackage{pdfpages}
\usepackage{mathrsfs}
\usepackage[scr=rsfs,cal=boondox]
{mathalfa}

\title{
How causal perspectives can inform neuroscience data analysis
}

\author{
    Eric W.~Bridgeford$^{1,\dagger}$,
    Brian Caffo$^2$,
    Maya B. Mathur$^{1,*}$,
    Russell A. Poldrack$^{1,*}$.
    \thanks{
     $^1$ Stanford University, $^2$ Johns Hopkins University, $^*$ these authors share senior authorship,
    $^\dagger$ Corresponding author:
      Eric W.~Bridgeford (\email{ericwb@stanford.edu}).}
}

\makeatletter
\def\thanks#1{\protected@xdef\@thanks{\@thanks
        \protect\footnotetext{#1}}}
\makeatother
\makeatletter
\renewcommand\@biblabel[1]{#1.}
\makeatother
\begin{document}

\maketitle

\input{Content/abstract}
\input{Content/intro}

\input{Content/basics}
\input{Content/confounders}
\input{Content/colliders}

\input{Content/discussion}
\input{Content/other_stuff}

\bibliographystyle{unsrtnat}
\bibliography{causal_pers}

\newpage

\appendix
\input{Content/appendix}

\end{document}

%% file: sf_preamble.tex
\usepackage{comment}
\usepackage{lipsum}
\usepackage[margin=1in, footskip=36pt]{geometry}

\usepackage{endnotes}

\usepackage{titlesec}
\usepackage{titletoc}
\usepackage{pgfplotstable}
\titleclass{\task}{straight}[\section]
\newcounter{task}
\renewcommand{\thetask}{\arabic{task}}
\titleformat{\task}[hang]
    {\normalfont\LARGE\bfseries}{Task \thetask:}{1em}{}

\titleformat*{\task}{\color{header1}\bfseries}

\titlecontents{task}
              [3.8em] 
              {}
              {\contentslabel{2.3em}}
              {\hspace*{-2.3em}}
              {\titlerule*[1pc]{.}\contentspage}

\titlespacing*{\section}{0ex}{1ex}{1ex}
\titlespacing*{\subsection}{0ex}{1ex}{1ex}
\titlespacing*{\subsubsection}{0ex}{1ex}{1ex}
\titlespacing*{\paragraph}{0ex}{1ex}{1ex}
\titlespacing*{\subparagraph}{0pt}{1ex}{1ex}
\titlespacing*{\task}{0em}{1ex}{1ex}

\usepackage[sort&compress,comma,square,numbers]{natbib}

\usepackage{amsfonts,amsmath,amssymb}
\usepackage{bbm}

\usepackage{multicol}
\usepackage{enumitem}
\setlist[enumerate]{wide, labelindent=1cm,  noitemsep}
\setlist[itemize]{noitemsep}
\setlist[description]{noitemsep}

\usepackage{hhline}

\usepackage{todonotes}
\RequirePackage{ulem}

\definecolor{tableheadcolor}{gray}{0.92}
\newcommand{\topline}{ %
        \arrayrulecolor{maroon}\specialrule{0.1em}{\abovetopsep}{0pt}%
        \arrayrulecolor{tableheadcolor}\specialrule{\belowrulesep}{0pt}{0pt}%
        \arrayrulecolor{maroon}}
\newcommand{\midtopline}{ %
        \arrayrulecolor{tableheadcolor}\specialrule{\aboverulesep}{0pt}{0pt}%
        \arrayrulecolor{maroon}\specialrule{\lightrulewidth}{0pt}{0pt}%
        \arrayrulecolor{white}\specialrule{\belowrulesep}{0pt}{0pt}%
        \arrayrulecolor{maroon}}
\newcommand{\bottomline}{ %
        \arrayrulecolor{white}\specialrule{\aboverulesep}{0pt}{0pt}%
        \arrayrulecolor{maroon} %
        \specialrule{\heavyrulewidth}{0pt}{\belowbottomsep}}%

\pgfplotstableset{normal/.style ={%
        header=true,
        string type,
        font=\small,
    	column type={p{.092\textwidth}},
        every head row/.style={
            before row={\topline\rowcolor{tableheadcolor}},
            after row={\midtopline}
        },
        every odd row/.style={
            before row={\rowcolor{maroon}}
        },
        every last row/.style={
            after row=\bottomline
        },
        col sep=&,
        row sep=\midrule
    }
}

\usepackage{multirow}

\usepackage{booktabs}
\usepackage{fancyhdr}
\usepackage{fullpage}

\RequirePackage{titlesec}
\RequirePackage[font={footnotesize},{singlelinecheck=false}]{caption}
\usepackage[T1]{fontenc}
\usepackage[scaled]{helvet}
\usepackage[scaled=1.10, med]{zlmtt}

\usepackage{algorithm}
\usepackage{algpseudocode}



%% file: Content/abstract.tex
\begin{abstract}
Over the past two decades, considerable strides have been made in advancing neuroscientific techniques, yet challenges remain in attributing causality to observed associations. This review addresses a fundamental issue in observational neuroscience studies and advocates for incorporating causal inference frameworks into standard practice. We systematically introduce necessary definitions and concepts, emphasizing how causal assumptions underlie statistical analyses even when not explicitly stated. Through a running example on sleep quality and white matter integrity, we illustrate how persistent challenges, including confounding and selection biases, can be conceptualized and addressed using causal frameworks. We demonstrate practical approaches for making assumption violations transparent through hands-on examples: supplementary case studies using multi-site harmonization and head motion exclusion procedures provide step-by-step diagnostic techniques for checking covariate overlap and identifying selection bias through exclusion pattern analysis. We explore how these causal perspectives can inform both experimental design and analytical choices, particularly for observational studies where traditional randomization is infeasible. Together, we believe this framework offers concrete tools for strengthening causal interpretations and inspiring more robust approaches to problems in neuroscience.
\end{abstract}

%% file: Content/intro.tex
\section*{Introduction}

Linear statistical models are a fundamental workhorse of neuroscience. Nearly every paper published in a neuroscience journal uses statistical methods such as analysis of variance (ANOVA), correlation analysis, or regression modeling to estimate associations between variables and test hypotheses. These methods are linear in the parameters, allowing for nonlinear relationships like age$^2$ or interactions, while remaining computationally tractable. As an example of their prominence, a review of articles published in a recent issue of \textit{Nature Neuroscience} (Volume 27 Issue 4, April 2024) found that 14 out of 15 articles presented results from ANOVA (12 articles) and/or regression modeling (8 articles), demonstrating how central these approaches are to drawing scientific conclusions about brain function.

One of the slogans taught in nearly every introductory statistical class is that ``correlation does not imply causation.'' However, associations identified using statistical models are almost always interpreted as reflecting underlying causal mechanisms to some degree. This interpretation is scientifically necessary because the description of causal mechanisms is central to scientific explanation \cite{Bechtel2005Jun}. A fundamental insight of causal inference research is that the valid interpretation of any statistical model relies upon an understanding of the causal relations between the variables in the model as well as other variables not included in the model, \textbf{even if the goal is not to develop an explicitly causal or mechanistic model}.

We propose that neuroscience should more formally incorporate the insights that have been developed over the last three decades within the field of causal inference. This review provides a ground-up treatment that bridges causal inference concepts with neuroscience practice, using neuroimaging examples to illustrate key principles. Rather than replacing existing methods, this framework helps researchers specify the conditions under which their chosen statistical approach can yield reliable and valid inferences. When these assumptions are met, simple statistical methods can provide robust insights; when violated, even sophisticated analyses may mislead.

This review complements recent neuroscience work on causation. \citet{Ross2024Feb} distinguish mechanistic from statistical causation, while \citet{Reid2019Nov} review causal methods for connectivity analysis. Building on this foundation, we focus on the practical challenge of determining when standard statistical approaches support valid causal interpretations. Recent work has also highlighted how population diversity and study design choices affect the generalizability of neuroimaging conclusions \cite{osayande2025diversity,kopal2023end}, underscoring the importance of explicit assumptions about when and how statistical associations reflect causal relationships. We demonstrate these identification principles through the example of sleep quality effects on white matter integrity, while supplementary sections provide advanced material on specialized scenarios including multi-site harmonization and head-motion exclusion procedures. Although we focus on neuroimaging examples, these identification principles and data patterns that emerge apply broadly in observational research in psychology, cognitive science, and related fields.

%% file: Content/basics.tex
\section{Identifying variables and estimands of interest}
\label{sec:s1}

Throughout this paper we will use a running example of the relationship between sleep quality and white matter structure.   Neuroscience evidence suggests that aspects of brain structure can be associated with lifestyle factors, including sleep quality. Poor sleep has been linked to accelerated brain aging and cognitive decline \cite{Mander2017,Sexton2014}, making it a target for potential interventions. Longitudinal studies suggest that sleep disturbances precede and may contribute to white matter deterioration, rather than the reverse \cite{Fjell2014,Sexton2014}. Sleep deprivation has been shown to disrupt glymphatic clearance of metabolic waste from the brain, potentially leading to white matter damage over time \cite{Xie2013}. In causal inference, it is useful if the factor being studied can be experimentally manipulated in the real world \cite{Hill1965May}.  In this example, sleep can be changed through behavioral interventions, sleep restriction protocols, or sleep enhancement techniques.

To illustrate this, researchers could study the effect of randomly assigning individuals to either (1) a group where sleep is regularly interrupted throughout the night, or (2) a control group provided optimal sleeping conditions. Measures of white matter integrity collected before and after the intervention would measures changes related to the manipulation,  and the difference in outcomes between the two groups would quantify the causal effect of the sleep intervention.

\begin{tcolorbox}[colback=blue!10,colframe=blue!50,title=Working Example: Sleep Quality and White Matter Integrity,breakable]
\textbf{Research Question}: Does poor sleep quality cause reduced white matter integrity in older adults?\\
\textbf{Exposure}: Sleep condition group (interrupted or control) \\
\textbf{Outcome}: White matter integrity (measured via diffusion weighted imaging, DWI, fractional anisotropy)\\
\textbf{Causal Estimand}: Average Treatment Effect (ATE), the average difference in white matter integrity if everyone had ideal sleep quality versus if everyone had interrupted sleep \\
\textbf{Why This Matters}: Understanding whether sleep interventions could preserve brain structure has implications for preventing cognitive decline
\end{tcolorbox}

\subsection{The basics of causal inference and causal graphs}

\begin{figure}
    \centering
    \includegraphics[width=\linewidth]{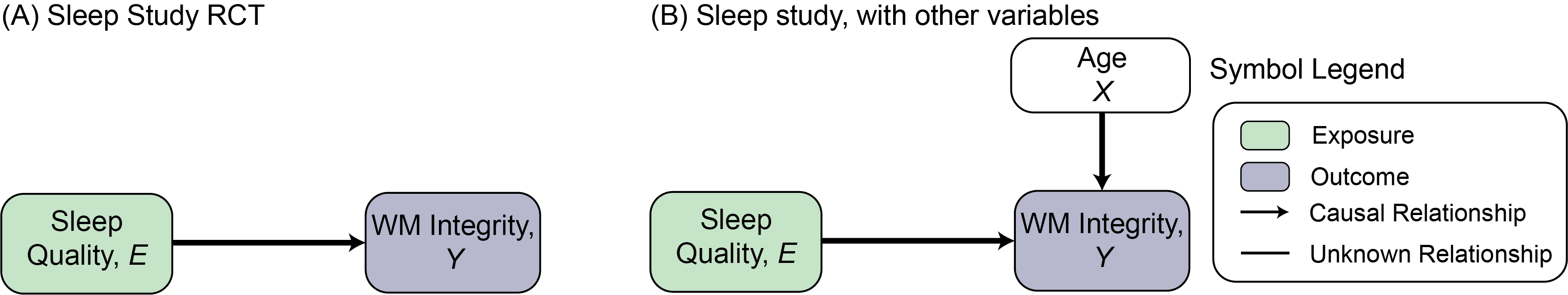}
    \caption{\textbf{Directed Acyclic Graphs (DAGs) encode causal relationships.} \textbf{(A)} a simple sleep quality example, where Sleep Quality is hypothesized to causally affect White Matter Integrity. \textbf{(B)} an additional variable, Age, also affects White Matter Integrity. However, due to randomization of sleep quality group, Age does not have an effect on Sleep Quality intervention group.}
    \label{fig:simple_dags}
\end{figure}

Causal inference typically begins by defining variables under study, which summarize characteristics being analyzed as well as factors that affect these characteristics. In our sleep example, variables include the experimentally manipulated variable (often called the ``exposure'', Sleep Quality in our example) and the outcome (White Matter Integrity). Causal relationships between variables are denoted by arrows, which encode causal mechanisms. For example, in Figure \ref{fig:simple_dags}(A), Sleep Quality $\rightarrow$ White Matter Integrity encodes the causal claim that sleep quality causes changes in white matter integrity. Here, we avoid a complexity by equating {actual} sleep quality and randomization group; for example some randomized to no sleep interruptions may have poor sleep quality for some other reason. Analyzing the randomization group as if it were the actual exposure, rather than a proxy for it, utilizes the well accepted intent to treat (ITT) principle \cite{gupta2011intention,fisher1990intention}.

For a particular causal relationship, the variable at the base of an arrow is the exposure (``the cause,'' typically denoted by $E$), while the variable at the head is the outcome (``the affected variable,'' typically denoted by $Y$). This requires that the outcome occur temporally after the exposure. For instance, if we implement a sleep intervention, this may cause an individual's white matter integrity to change, whereas changes in white matter integrity would not cause assignment to a sleep intervention.

Usually, an analysis focuses on one specific relationship, known as the estimand of interest: the relationship that one wishes to estimate and study. A common causal estimand is the average treatment effect (ATE), which represents the average effect that an exposure has on an outcome. In our example, the ATE quantifies the average difference in white matter integrity if everyone had good sleep quality versus poor sleep quality. We learn about this estimand through the estimator: a rule for estimating a desired effect from observed data.

\paragraph{Identification of causal effects} The goal for causal analyses is to determine how the outcome would change for a given intervention on the exposure. In our working example, we wish to clarify how white matter integrity changes if an individual were made to have good sleep quality versus poor sleep quality. This differs from simply asking how outcomes differ across naturally occurring exposure groups; causal questions focus on the actual impact of the exposure itself, rather than just whether exposure groups differ.

Even in idealized experimental neuroscience studies, this is not directly estimable due to the fundamental problem of causal inference: at any given single time, we can only observe each individual under one exposure condition, and never both. For any given person, at any given time, we can observe their white matter integrity with either good or poor sleep quality. We could potentially measure both outcomes for the same individual via crossover studies, but these outcomes would occur at different times \cite{sibbald1998understanding}. This means that without additional assumptions, we cannot directly calculate individual-level causal effects. Causal inference is therefore concerned with identification (ID) assumptions: the assumptions under which conclusions about the causal effect can be drawn using observed data despite this fundamental limitation. ID assumptions define whether there exists any estimator that could validly estimate the causal effect of interest by comparing across different individuals who received different exposures.

When we say that a technique provides evidence of a causal relationship, we mean that if enough data were collected with appropriate assumptions, the technique will correctly estimate the true underlying estimand of interest (which may be causal or non-causal); this is referred to as statistical consistency. Statistical consistency is closely related to identification. Identification asks whether a causal effect of interest can in principle be written as a function of observed data, while consistency asks whether our specific estimation technique (e.g., ANOVA) can recover a particular effect (which may or may not be causal) given sufficient data.

For instance, identification would ask whether it is theoretically possible to recover the causal effect of sleep quality on white matter integrity; consistency would ask whether ANOVA would recover the effect with enough data. When data do not meet required assumptions but we still attempt to draw causal conclusions, our conclusions can be biased, meaning that they systematically mis-estimate the underlying effect. This bias could be due to a failure of identification (i.e., the assumptions we understand about the data do not permit causal inferences) or consistency (i.e., our chosen estimation technique improperly estimates our desired effect, which may or may not be causal). The relationship between consistency and identification is further clarified in Table \ref{tab:cmp_id_cons}. 

\begin{table}[h]
\begin{tabular}{cccp{9cm}}
\toprule
\textbf{Identified?} & \textbf{Consistent?} & \textbf{Result} & \textbf{What this means} \\
\midrule
\multicolumn{4}{l}{\textit{For association/correlation questions:}} \\
-- & $\checkmark$ & \textcolor{green!70!black}{Valid} & Associations generally depend on observable aspects of the data data, so are generally identifiable; just need right method \\
-- & $\times$ & \textcolor{red}{Biased} & Wrong method gives wrong association \\
\midrule
\multicolumn{4}{l}{\textit{For causal questions:}} \\
\checkmark & \checkmark & \textcolor{green!70!black}{Valid} & The causal effect exists in the data and your method finds it correctly \\
\checkmark & $\times$ & \textcolor{red}{Biased} & The answer could be obtained but your method gets it wrong (e.g., using linear regression for nonlinear relationships) \\
$\times$ & -- & \textcolor{red}{Biased} & The answer is not in the data -- no method can find it \\
\bottomrule
\end{tabular}
\caption{\textbf{The relationship between causal identification, statistical consistency, and valid inference.} For association questions (e.g., ``Do sleep groups differ in white matter integrity?''), statistical consistency is most relevant; associations typically only depend on observable aspects in data, so generally additional assumptions are needed not regarding identification as long as the observed data contains those quantities. For causal questions (e.g., ``Does poor sleep cause reduced white matter integrity?''), both causal identification and statistical consistency are required. When causal effects are not identified (such as in observational studies with unmeasured confounders, discussed in Section \ref{sec:confounding}), no statistical method can consistently estimate them, making valid causal inference impossible.}
    \label{tab:cmp_id_cons}
\end{table}

\paragraph*{Randomized experiments}

A randomized experiment, often called a randomized controlled trial (RCT) in the context of medicine, is a study in which individuals are randomly assigned to exposure conditions. When data are collected under randomized design, the exposure assignment is not determined by other variables: it is imposed by the assignment mechanism (such as a random number generator) rather than being influenced by participant characteristics.

Randomization is sufficient for causal identification because with high probability it ensures that exposure groups are comparable (on average) across all variables, known and unknown, measured and unmeasured \cite{rubin1974estimating}. Each person has an equal chance of being assigned to either sleep group regardless of any other characteristics (e.g., their age, stress levels, or health status), and the groups will be similar with respect to all of these other potential factors, on average. The only systematic difference between the groups is the sleep intervention group itself. Differences in outcomes between groups can therefore be attributed to the exposure itself rather than to other factors.

When the exposure is randomized, unadjusted bivariate analyses provide evidence of causal relationships with minimal additional assumptions \cite{Hill1965May}. An unadjusted bivariate analysis examines the relationship between an exposure and an outcome without controlling for other variables. For example, one might use ANOVA to compare white matter integrity across sleep quality groups, or correlation analysis to assess the association between sleep quality and white matter measures. In the sleep example, such an analysis of white matter integrity with sleep quality group is sufficient to derive causal conclusions using these simple methods because sleep quality group is randomized across individuals. Consider, for instance, that age is related to white matter integrity, in Figure \ref{fig:simple_dags}(B). Due to the randomized assignment of individuals across sleep groups, age is not causally related to the sleep intervention group (no arrow from Age $\rightarrow$ Sleep Quality). When other variables are unrelated to the exposure, unadjusted bivariate analyses permit causal inferences for the effect of the exposure on the outcome (i.e., causal effects can be identified and unadjusted bivariate analyses can consistently estimate them) with few additional assumptions.

\begin{tcolorbox}[colback=gray!10,colframe=gray!50,title=Key Causal Terminology,breakable]
\textbf{Variables}: Characteristics being analyzed, including factors that affect these characteristics\\
\textbf{Exposure}: The presumed cause in a causal relationship\\
\textbf{Outcome}: The presumed effect in a causal relationship\\
\textbf{Directed arrows}: Relationships showing cause-and-effect (e.g., $E \rightarrow Y$)\\
\textbf{Causal graph (DAG)}: Visual representation of assumed causal relationships between variables\\
\textbf{Fundamental problem of causal inference}: Inability to observe individuals under all possible exposure conditions\\
\textbf{Average Treatment Effect (ATE)}: Average difference in outcomes under different exposure levels\\\textbf{Statistical Consistency}: Whether a chosen statistical method correctly estimates an effect (causal or non-causal)\\
\textbf{Identification}: Whether a causal effect can theoretically be estimated from observed data
\end{tcolorbox}

\paragraph*{The inadequacy of bivariate association in nonrandomized studies}

Unadjusted bivariate analysis is often insufficient on its own for deriving causal conclusions in observational studies. An observational study is a study in which researchers do not have direct control of the exposure, intervention, or cause of interest (even if they hypothetically could) \cite{hernan2020causal}. Observed associations between the exposure and outcome may be attributable to genuine causal effects or to biases induced by failures to appropriately handle additional variables \cite{Ciolino2013,Power2017Jan}.

When the exposure and outcome are related in ways that are not causal, unadjusted bivariate analyses can fail to distinguish genuine causal effects from these other sources of association. Unadjusted bivariate analyses can be subject to biases yielding spurious conclusions, including erroneously detected effects (false positives), mistakenly non-detected effects (false negatives), and other ways that conclusions may not generalize outside the specific study sample collected.

%% file: Content/confounders.tex
\section{Identifying and controlling for confounding bias}
\label{sec:confounding}

In the previous section, we saw how randomized experiments enable straightforward causal inference through simple statistical methods. However, many neuroscience studies (particularly in the context of ``population neuroimaging'') rely on observational data where the exposure of interest is not under experimental control. In these settings, multiple regression with covariates has become standard practice for estimating relationships while controlling for potential covariates like age, sex, and many other ``nuisance'' variables. While this represents theoretically sound statistical methodology, the validity of causal interpretations depends on identification (ID) assumptions that may be violated in ways that are not immediately apparent from standard analyses.

This section examines when and why standard regression approaches succeed or fail for causal inference, building on the concept of confounding bias and methods to address it \cite{Pearl2009Sep,hernan2020causal}. We continue with our sleep quality and white matter integrity example, now considering it in an observational context where participants naturally vary in their sleep patterns rather than being randomly assigned to experimental conditions.

\begin{tcolorbox}[colback=blue!10,colframe=blue!50,title=Working Example: Observational Sleep Study,breakable]
\textbf{Research Question}: Does poor sleep quality cause reduced white matter integrity in older adults?\\
\textbf{Study Design}: Cross-sectional study of 200 adults aged 50-80 recruited from community and memory clinic\\
\textbf{Exposure}: Sleep quality (Pittsburgh Sleep Quality Index scores)\\
\textbf{Outcome}: White matter integrity (DWI fractional anisotropy)\\
\textbf{Key common cause of exposure and outcome}: Age (affects both sleep patterns and brain aging)\\
\textbf{Challenge}: Older participants tend to have worse sleep AND lower white matter integrity due to normal aging: how do we separate sleep effects from age effects?
\end{tcolorbox}

\subsection{Understanding confounding through causal graphs}

Consider our sleep quality example in an observational setting where participants are not randomly assigned to good or poor sleep conditions. Instead, their sleep patterns emerge from complex interactions of age, lifestyle, health status, and environmental factors. This creates the potential for confounding bias: spurious associations between sleep and brain structure that arise not from direct causal relationships, but from common causes of both variables.

\begin{figure}
    \centering
    \includegraphics[width=\linewidth]{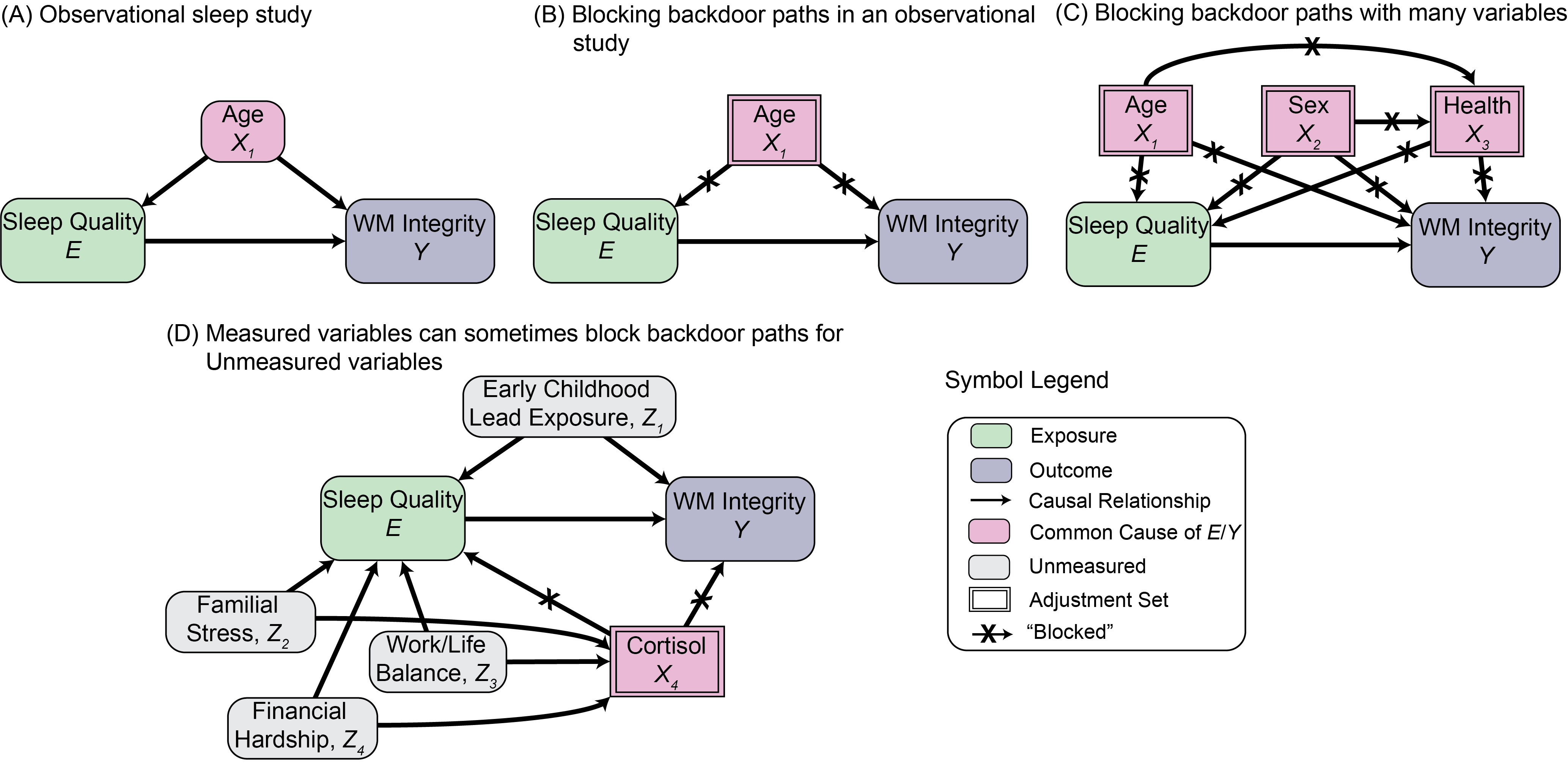}
    \caption{\textbf{Confounding in sleep-brain relationships}. \textbf{(A)} Age creates a backdoor path between sleep quality and white matter integrity. \textbf{(B)} Controlling for age (indicated by the square) blocks the backdoor path, isolating the effect of the exposure (Sleep Quality) on the outcome (WM Integrity). \textbf{(C)} Multiple common causes (age, sex, health status) create multiple backdoor paths that must all be blocked. \textbf{(D)} certain unmeasured variables can be blocked when the link between the exposure or the outcome can be blocked by a measured variable (e.g., familial, work/life, or financial stressors, blocked by conditioning on cortisol levels). Other unmeasured variables (e.g., lead exposure during childhood) may not have this advantage.}
    \label{fig:sleep_confounding}
\end{figure}

Figure \ref{fig:sleep_confounding}(A) illustrates a simple confounding scenario. Age is a common cause of both sleep quality and white matter integrity. Older adults tend to experience more sleep disruption due to circadian rhythm changes, medical conditions, and medication effects, while simultaneously showing natural age-related decline in white matter integrity. Without accounting for age, an analysis might incorrectly attribute normal age-related brain changes to poor sleep quality.

This confounding manifests as a {backdoor path}: Sleep Quality $\leftarrow$ Age $\rightarrow$ White Matter Integrity, which is a path that starts with an arrow pointing into the exposure (Sleep Quality $\leftarrow$ Age) and ends with an arrow pointing into or out of the outcome (Age $\rightarrow$ White Matter Integrity). This contrasts from a causal path, which is a path from the exposure to the outcome where all of the arrows point in the same direction towards the outcome. Unlike the direct causal path of interest here (Sleep Quality $\rightarrow$ White Matter Integrity), this backdoor path represents a non-causal (i.e., the arrows are not all pointing from the exposure to the outcome) source of association between the exposure and the outcome\footnote{Note that the presence of causal and backdoor paths depends on the research question of interest. For instance, if we were studying the effect of aging on WM integrity, under the exact same DAG, there would be a direct causal effect of Age $\rightarrow$ WM integrity, and an indirect (but still causal) effect of Age $\rightarrow$ Sleep Quality $\rightarrow$ WM Integrity.}. When backdoor paths are present, simple correlations or unadjusted comparisons conflate causal effects of interest (Sleep Quality on WM Integrity) with non-causal associations (due to the backdoor path Sleep Quality $\leftarrow$ Age $\rightarrow$ White Matter Integrity). This can lead to misleading conclusions about sleep interventions or their clinical relevance.

The goal of confounding control is to ``block'' these backdoor paths while preserving the causal pathway of interest \cite{pearl1995causal,Greenland1996May}. When successful, this isolates the direct causal effect of sleep on white matter integrity from confounding influences. However, achieving this goal requires meeting several stringent assumptions.

\subsection{Core identification assumptions}

Valid causal inference from observational data requires four core ID assumptions \cite{hernan2020causal,Imbens2015Apr}. Understanding these assumptions, and when they might be violated, is essential for interpreting observational studies that make (explicitly or implicitly, through interpretation or generalization) causal claims. Unlike statistical consistency assumptions that focus on whether a particular estimator recovers the true parameter, ID assumptions concern whether any estimator could theoretically identify causal effects from the observed data.

\paragraph{Conditional ignorability (No unmeasured confounding)} requires that we appropriately control for variables that influence both exposure and outcome using measured covariates \cite{rosenbaum1983central,Imbens2015Apr}. In formal terms, the measured covariates capture the systematic reasons why individuals in different exposure groups would differ in their outcomes other than the exposure itself. For our sleep example, this means we have identified and controlled for factors that influence both sleep quality and white matter integrity.

This assumption is often the most challenging to satisfy in practice. Genetic factors might influence both sleep architecture and brain structure through shared neurobiological pathways \cite{Sehgal2011,Dashti2019}. Socioeconomic status might affect both sleep environment (via e.g. noise, stress, shift work) and brain health (via e.g. diet, health care availability, or environmental exposures) \cite{Grandner2022,Shaked2022}. Comorbid conditions like depression or diabetes might simultaneously disrupt sleep and contribute to white matter pathology \cite{wang2020relbtwn,Hollocks2015}. If any such factors remain unmeasured, conditional ignorability can be violated and standard confounding control methods may be insufficient.

The assumption becomes more plausible when researchers can identify and measure the most important common causes based on domain knowledge, prior research, and biological understanding. However, it can never be verified from data alone, as there is always potential for unmeasured confounding to affect conclusions.

\paragraph{Positivity (Overlap)} requires that individuals with any given set of characteristics could plausibly have received any exposure level \cite{rubin1974estimating}. In other words, for any combination of participant characteristics (age, health status, etc.), it should be plausible that there are some people with good sleep and some with poor sleep.

In sleep studies, positivity would be violated if certain combinations of characteristics were associated with only good or only poor sleep. For example, if all young, healthy participants had good sleep while all older participants with medical conditions had poor sleep, we could not separate age and health effects from sleep effects because the covariate distributions would not overlap sufficiently across exposure groups to support causal inference.

Positivity violations are particularly common in observational studies that recruit from convenience samples \cite{Andrade2021,LeWinn2017}. University-based studies might oversample young, educated participants with good sleep, while clinical samples might include primarily older participants with sleep disorders. Such sampling patterns create fundamental limitations in the causal effects that can be identified from the observed data.

\paragraph{Consistency (Well-defined interventions)} requires that each level of the exposure corresponds to a single, well-defined intervention within the context of our population \cite{cole2009consistency,hernan2020causal}. The exposure variable must represent a coherent treatment or intervention that could be manipulated for the individuals under study.

For sleep quality, this assumption faces several challenges. ``Poor sleep'' might encompass different underlying ``interventions'': sleep restriction (reducing total sleep time to 4-5 hours), sleep fragmentation (frequent interruptions throughout the night), or circadian disruption (shifting sleep timing). Each represents a distinct intervention that could have different causal effects on white matter integrity.

If different participants classified as having ``poor sleep'' actually experienced different types of sleep interventions, the consistency assumption would be violated and the causal estimand becomes ill-defined. For consistency to hold, researchers must specify which particular sleep intervention their exposure represents; for example, ``sleep restriction to $\leq 5$ hours per night'' rather than the broader category of ``poor sleep quality.''

\paragraph{No interference} requires that each individual's outcome depends only on their own exposure, not on other participants' exposures \cite{rubin1980randomization,rubin1986comment,Imbens2015Apr}. This assumption rules out spillover effects between study participants.

In sleep studies, interference might occur if participants are couples, family members, or roommates whose sleep patterns influence each other. Intervention studies might result in spillover if participants share information about sleep hygiene strategies.

Together, the no interference and consistency assumptions are collectively known as the Stable Unit Treatment Value Assumption (SUTVA) \cite{rubin1980randomization,rubin1986comment,Imbens2015Apr}.

\begin{tcolorbox}[colback=gray!10,colframe=gray!50,title=Key Confounding Concepts,breakable]
\textbf{Common cause (confounder)}: Variable that affects both exposure and outcome (Age affects both Sleep Quality and White Matter Integrity)\\
\textbf{Causal path}: a pathway from an exposure to an outcome where all of the arrows point in the same direction \\
\textbf{Backdoor path}: Non-causal pathway through common causes that points into the exposure (Sleep Quality $\leftarrow$ Age $\rightarrow$ White Matter Integrity)\\
\textbf{Blocking}: Statistical conditioning that closes backdoor paths\\
\textbf{Adjustment set}: A conditioning set of measured variables which closes backdoor paths\\
\textbf{Conditional ignorability}: All common causes are controlled for\\
\textbf{Positivity}: All covariate patterns exist across exposure groups\\
\textbf{Consistency}: Well-defined interventions with stable effects\\
\textbf{No interference}: Outcomes for each individual depend only on their exposure\\
\textbf{SUTVA}: A blanket term for the consistency and no interference assumptions
\end{tcolorbox}

\subsection{Methods for controlling confounding}
\label{sec:control_confound}
In general, a core issue in controlling confounding is the need to identify and measure a set of variables, known as the {adjustment set}, which block all backdoor paths. This is typically notated by adding square boxes around the variables being adjusted for, as indicated in Figure \ref{fig:sleep_confounding}(B) and Figure \ref{fig:sleep_confounding}(C). The X's are used to indicate that these specific causal relationships are blocked by conditioning on the adjustment variables (typically, these arrows would simply be deleted from the DAG). With these arrows blocked, the backdoor paths can no longer transmit non-causal associations between the exposure and the outcome, and the backdoor path is said to be closed. A set of all common causes of the exposure and outcome always forms a sufficient adjustment set to block all backdoor paths, when such variables are observed.

``Blocking'' of backdoor paths is the graphical equivalent to satisfying causal identification assumptions \cite{pearl1995causal}. When these assumptions are satisfied, several analytical approaches can identify causal effects from observational data \cite{Stuart2010Feb,hernan2020causal}. These methods differ in their robustness to assumption violations and their transparency about when assumptions fail.

\paragraph{Multiple regression} is the most commonly used approach in neuroimaging for controlling confounding. An \textit{outcome model} specifies how the outcome is influenced by the exposure and covariates \cite{Agresti2015Feb}; for example, modeling white matter integrity as a function of sleep quality, age, sex, and health status. This approach works well when relationships are approximately linear and all important common causes are measured.

However, multiple regression can mask positivity violations by extrapolating to covariate combinations that do not exist in the data. Consider a sleep study where participants with poor sleep are predominantly older adults recruited from memory clinics, while participants with good sleep are predominantly younger adults from university settings. A regression model would estimate sleep effects for young adults with poor sleep, even if no such participants exist in the data. The regression estimate for this combination relies entirely on extrapolation, assuming the functional form generalizes across age groups. Standard regression output provides no indication that estimates are based on extrapolation rather than observed data, potentially leading to unreliable conclusions about interventions in unrepresented populations \cite{Antonakis2014May}.

To address some of these limitations, researchers have developed \textit{positivity-aware methods} that make violations of ID assumptions more transparent; particularly, violations of the positivity assumption. Below, we describe a few of these techniques.

\paragraph{Stratification} divides participants into covariate-defined subgroups and estimates sleep effects within each stratum \cite{Jepsen2004Aug, Rothman2012Dec}. Results are then combined across strata using appropriate weights. This approach makes positivity violations transparent, as empty or sparse strata immediately reveal where causal effects cannot be reliably estimated.

However, stratification becomes impractical with many continuous covariates or high-dimensional covariate spaces, as the number of required strata grows exponentially \cite{Gelman2006Dec}. The approach works best with a small number of categorical or discretized common causes.

\paragraph{Propensity score methods} summarize all covariates into a single ``propensity score'' representing the probability of having poor sleep given an individual's characteristics \cite{rosenbaum1983central}. This dimension reduction makes complex confounding more manageable by reducing the covariate space to a single dimension (the propensity score).

The simplest approach is propensity trimming, which discards samples with propensity scores that are highly dissimilar from those in other exposure groups, reducing the number of samples with extreme propensities \cite{Robins1986Jan}. Since propensity scores encapsulate covariate information, this also removes samples with atypical covariate profiles that could otherwise bias subsequent analyses.

\textit{Inverse probability weighting} (IPW) assigns weights inversely proportional to propensity scores, then incorporates these weights into subsequent analyses to approximately align propensity distributions across exposure groups \cite{Robins1986Jan}. This process up-weights observations whose actual exposure status was unlikely given their covariates, indirectly balancing covariate distributions due to the relationship between covariates and propensity scores.

\paragraph{Matching methods} take a more direct approach by pairing samples from different exposure groups based on propensity or covariate similarity \cite{Stuart2008,Stuart2010Feb}. These methods aim to achieve covariate balance by pairing participants with similar characteristics but different sleep patterns. Samples without suitable matches are discarded from analysis, making sample limitations for the question of interest explicit and transparent.

Distance-based matching can offer greater flexibility than propensity-based approaches, as they do not require summarizing covariates into a single score. However, they require careful selection of distance metrics and can face challenges with high-dimensional covariate spaces \cite{Stuart2010Feb}.

These positivity-aware methods make assumption violations transparent through failed matches, empty strata, or extreme weights. However, they may substantially reduce effective sample sizes when covariate overlap is poor.

\subsection{Checking assumptions in practice}

It is important to note that the ID assumptions cannot be fully verified empirically. However, researchers can take concrete steps to help evaluate whether their analysis satisfies ID assumptions and to make assumption violations transparent. A hands-on example of exploratory approaches for assessing conditional ignorability and positivity is provided in Appendix \ref{sec:batch_effect}:

\paragraph{Evaluating conditional ignorability:}
\begin{enumerate}[leftmargin=*]
\item Construct a causal graph incorporating domain knowledge and prior research
\item Identify all potential backdoor paths between exposure and outcome. Consider leveraging tools such as \href{https://www.dagitty.net/}{DAGitty}, a web browser utility and R package \cite{Textor2016}, which allows users to input causal graphs and verify that measured covariates form proper adjustment sets
\item Consider sensitivity analysis for potential unmeasured common causes
\item Check balance of covariates across exposure groups after adjustment
\end{enumerate}

\paragraph{Assessing positivity:}
\begin{enumerate}[leftmargin=*]
\item Create overlapping histograms of covariate distributions by exposure group
\item Examine scatterplots of key covariates colored by exposure status  
\item Calculate propensity scores and identify extreme values (near 0 or 1)
\item For matching approaches: Report the proportion of participants who lack suitable matches
\item Create cross-tabulations of covariates of interest to identify empty cells for categorical variables
\item Consider using or adapting propensity-aware methods, where possible
\end{enumerate}

\paragraph{Evaluating consistency and SUTVA:}
\begin{enumerate}[leftmargin=*]
\item Clearly define what intervention or cause the exposure represents
\item Verify (via domain expertise with dataset collection protocols) that this exposure captures a similar phenomena across participants
\item Check for potential interference between participants
\item Consider whether different measurement methods might capture different constructs
\end{enumerate}

\begin{tcolorbox}[colback=blue!10,colframe=blue!50,title=Example: Multi-site Sleep Study with Positivity Violation,breakable]
\textbf{Study Design}: Sleep effects on white matter studied at university hospital (younger participants) and memory clinic (older participants with cognitive concerns)\\
\textbf{Data Pattern}: University site: 95\% good sleepers, mostly age 20-40. Memory clinic: 80\% poor sleepers, mostly age 60-80\\
\textbf{Positivity Violation}: No young participants with poor sleep, no older participants with good sleep\\
\textbf{Implication}: Cannot separate sleep effects from age effects or site effects\\
\textbf{Potential Solutions}: (1) Recruit across age ranges at both sites, (2) Restrict analysis to overlapping age range (40-60), (3) Acknowledge limitation and avoid causal claims\\
\textbf{Key Insight}: Causal identification requires sufficient overlap in participant characteristics across exposure groups
\end{tcolorbox}

\subsection{Unmeasured confounding and sensitivity analysis}

Even with careful design and analysis, unmeasured confounding remains a persistent threat to causal inference \cite{Greenland2005Mar,VanderWeele2017Aug}. In sleep research, numerous factors might influence both sleep patterns and brain structure while remaining unmeasured in typical neuroimaging studies. Consider two distinct types of unmeasured common causes illustrated in Figure \ref{fig:sleep_confounding}(D).

Stress-related unmeasured common causes might include work/life balance, financial hardship, and family stress. These factors operate primarily through cortisol, the key stress hormone affecting both sleep and white matter integrity \cite{cox2016white}. As shown in Figure \ref{fig:sleep_confounding}(D), conditioning on measured cortisol levels can block the backdoor paths from these unmeasured stressors because cortisol represents their common downstream mechanism.

Early developmental exposures like childhood lead exposure represent a different challenge. Lead exposure during critical periods permanently alters brain development through mechanisms that may leave no measurable adult biomarkers \cite{Needleman2004,Cecil2008}. As illustrated in Figure \ref{fig:sleep_confounding}(D), no measured variable can block this backdoor path because adult measurements may not capture the historical exposure that caused developmental effects.

This distinction illustrates why some unmeasured common causes can be addressed through proxy measurements while others cannot. When confounders operate through measurable biological mechanisms, controlling for those mechanisms provides protection. However, historical exposures with permanent developmental effects may not necessarily be captured by adult measurements, making sensitivity analysis essential for assessing robustness of causal conclusions.

\paragraph{Sensitivity analysis} provides a formal framework for assessing how robust conclusions are to potential unmeasured confounding \cite{Lin1998Sep,Liu2013Dec}. Rather than assuming no unmeasured common causes exist, sensitivity analysis asks: ``How strong would an unmeasured common cause need to be to explain away the observed association?''

For continuous outcomes, which are common in neuroimaging, the Robustness Value \citep{cinelli2020making} quantifies the minimum strength of association an unmeasured common cause would need with both exposure and outcome to reduce the estimated effect to zero. E-values \citep{VanderWeele2017Aug} provide a complementary approach for various outcome and exposure types (continuous or binary), and can be extended to survival outcomes.

These tools help researchers and readers evaluate the plausibility of unmeasured confounding explanations. A large robustness value or E-value suggests that only very strong unmeasured common causes could account for observed effects, making causal interpretations more credible. A small robustness value or E-value indicates vulnerability to even moderate unmeasured confounding, suggesting caution in causal interpretation.

\begin{tcolorbox}[colback=red!10,colframe=red!50,title=Section Takeaways,breakable]
\textbf{$\checkmark$ What works:} Multiple regression (or positivity aware methods) control confounding when all four ID assumptions are met and model is correctly specified\\
\textbf{$\times$ What fails:} Unmeasured common causes, positivity violations, poorly defined exposures, interference between units\\
\textbf{$\rightarrow$ What to do:} Draw causal graphs, check covariate overlap, define interventions clearly, consider sensitivity analysis for unmeasured confounding\\
\textbf{Key insight:} Standard adjustments (via regressions or positivity-aware methods) work well under specific assumptions; make these assumptions explicit and rational based on your domain expertise for the problem, and testable or visualizable where possible
\end{tcolorbox}

Understanding and addressing confounding represents a crucial first step toward valid causal inference in neuroimaging. However, confounding is not the only source of bias in observational studies. The next section examines how selection processes and data exclusion practices can introduce different types of bias that persist even when confounding is appropriately controlled.

%% file: Content/colliders.tex
\section{Selection bias and colliders}
\label{sec:selection}

The previous section examined confounding bias, where common causes create spurious associations between exposures and outcomes.  This section addresses a different but equally important threat to causal inference: selection bias. Selection bias occurs when the process of including or excluding participants from analysis introduces systematic differences that distort causal effect estimates. Unlike confounding, which involves variables that affect both exposure and outcome, selection bias often involves variables that are affected by both exposure and outcome, creating structures known as colliders.

\subsection{Understanding colliders through gallbladder disease and diabetes}

Consider a classic example from medical statistics known as Berkson's paradox \cite{berkson1946limitations}. In the general population, two medical conditions, cholecystitis (gallbladder disease) and diabetes, are generally unrelated. However, when analyzing only hospitalized patients, these conditions often appear positively associated, falsely suggesting one disease predisposes to the other.

This spurious association had serious clinical consequences. As Berkson noted, the apparent correlation led some medical practitioners to remove gallbladders as a treatment for diabetes, an invasive surgery based entirely on a statistical artifact. Patients underwent unnecessary operations because researchers failed to recognize how hospital-based sampling could create false associations.

\begin{figure}
    \centering
    \includegraphics[width=0.8\linewidth]{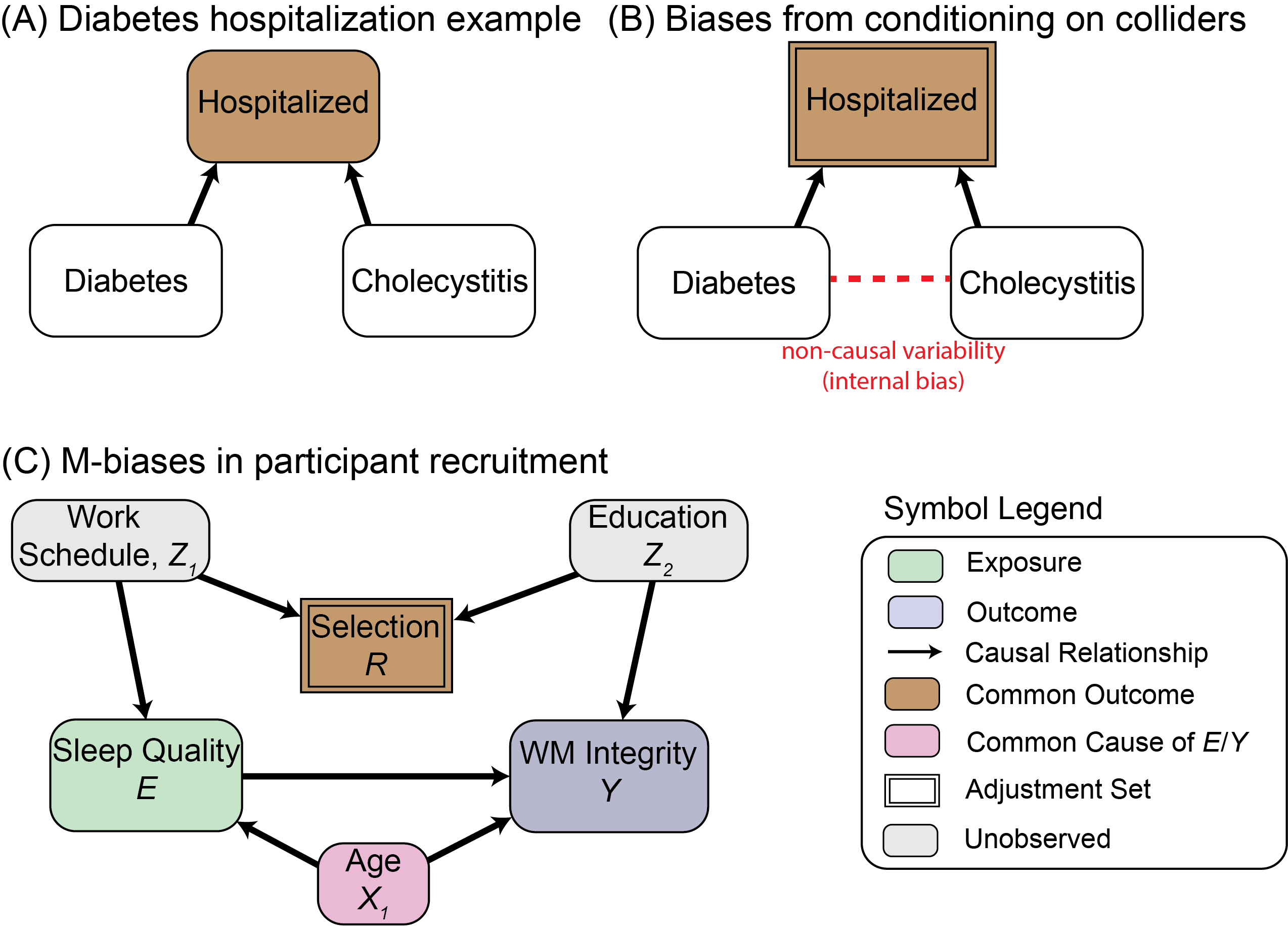}
    \caption{\textbf{Selection bias and collider structures}. \textbf{(A)} Berkson's paradox illustrating collider bias: both diabetes and cholecystitis independently increase hospitalization probability, but these conditions are unrelated in the general population. \textbf{(B)} Conditioning on hospital admission (indicated by the square) creates collider bias, introducing an artificial positive association between diabetes and cholecystitis (red dashed arrow) that does not exist in the broader population. \textbf{(C)} M-bias structure in sleep-brain research: two unmeasured variables (Work Schedule and Education Level) influence both the exposure-outcome relationship and study selection. Even when controlling for measured common causes like Age (indicated by the square), this creates spurious associations through the backdoor path: Sleep Quality $\leftarrow$ Work Schedule $\rightarrow$ [Selection] $\leftarrow$ Education Level $\rightarrow$ White Matter Integrity.}
    \label{fig:collider_examples}
\end{figure}

The mechanisms underlying this phenomena are relatively straightforward: both cholecytitis and diabetes affect hospitalization (Figure \ref{fig:collider_examples}(A)). These mechanisms are independent: diabetes leads some people to be hospitalized for conditions relating to blood sugar management, while cholecystitis leads others to be hospitalized for issues like gallbladder pain. In the general population, having one condition does not affect the probability of having the other.

In this example, hospital admission is a {collider} because it is a common outcome of two independent conditions: the two causal arrows ``collide'' at the hospitalization variable. When we restrict our analysis to hospitalized patients, shown by the square around hospital admission in Figure \ref{fig:collider_examples}(B), we introduce collider bias. This creates an artificial positive association between cholecystitis and diabetes (illustrated conceptually by the red dashed arrow) that does not exist in the broader population, and was interpreted by some pracititioners as a causal effect of cholecystitis on the development of diabetes.

Berkson demonstrated this empirically using hospital data. In his theoretical general population, diabetes and cholecystitis were completely independent; having one did not affect the probability of having the other (both diabetics and non-diabetics showed nearly equal cholecystitis rates). However, in his hospital population, where both conditions independently increase the chance of hospitalization, a spurious positive correlation emerged, with substantially more diabetics showing cholecystitis. Berkson found that this was an artifact of the relative frequencies of diseases in the hospital compared to the general population; the spurious positive correlation was an artifact of the selection process.

This illustrates a key principle: conditioning on variables like colliders can introduce bias rather than removing it. If we wanted to study whether diabetes causes some other outcome (like cognitive decline), restricting our analysis to hospitalized patients or conditioning on the covariate of hospitalization would bias our conclusions because we have, implicitly or explicitly, conditioned on a collider.

\subsection{Selection bias in neuroimaging: The M-bias problem}

Neuroimaging studies face similar challenges when participant recruitment creates collider structures. Consider our ongoing sleep quality example, but now focus on how participants end up in the study sample. Research studies rarely recruit random samples from the entire population; instead, recruitment is influenced by factors like work schedules and educational background.

\begin{tcolorbox}[colback=gray!10,colframe=gray!50,title=Key Collider Concepts,breakable]
\textbf{Common Outcome (Collider)}: A variable that is caused by two or more other variables\\
\textbf{Collider bias}: Spurious associations created by conditioning on a common outcome (collider) \\
\textbf{Selection bias}: A form of collider bias in which systematic differences in who gets included in sample can distort causal conclusions\\
\textbf{M-bias}: A specific collider configuration that creates bias even when controlling for measured common causes\\
\textbf{Internal bias}: When estimates in selected sample do not reflect true effects within that sample\footnotemark\\
\textbf{Net-external bias}: When causal effects in selected sample do not match those in target population\footnotemark[\value{footnote}]
\end{tcolorbox}
\footnotetext{These definitions apply when exposure does not affect selection. See \citet{Mathur2024Jun} for more complex cases where exposure affects selection.}

Figure \ref{fig:collider_examples}(C) illustrates this scenario using an M-bias structure \cite{Greenland2003,VanderWeele2012}. This structure is known as an M-bias because of the M-like shape created by the exposure $E$, unmeasured variables $Z_1$ and $Z_2$, the recruitment selection $R$, and the outcome $Y$. Two unmeasured variables influence the exposure our the outcome as well as study selection:
\begin{itemize}[leftmargin=*]
\item \textbf{Work Schedule} ($Z_1$): Shift workers or those with demanding schedules may have poor sleep quality \cite{Kecklund2016,Torquati2018} and may also be less likely to participate in research due to time constraints \cite{Biegus2022}
\item \textbf{Education Level} ($Z_2$): Higher education is associated with better white matter integrity through cognitive reserve mechanisms \cite{Stern2012,Barulli2013}, and educated individuals are more likely to participate in research studies \cite{Fry2017}
\end{itemize}

Both unmeasured variables influence Selection ($R$), creating a collider structure. Even though Age is measured and controlled for, this M-bias structure can introduce spurious associations between sleep quality and white matter integrity through the path: Sleep Quality $\leftarrow$ Work Schedule $\rightarrow$ [Selection] $\leftarrow$ Education Level $\rightarrow$ White Matter Integrity.

Importantly, \citet{Mathur2024Jun} demonstrate in Table 2, Row C that conditioning on the unmeasured variables that affect selection and the outcome (in this case, Education Level) would be sufficient to eliminate selection bias. However, if this variable is unmeasured, standard analytical approaches cannot eliminate the bias post-hoc and identify a causal effect free from internal and net-external bias.

\begin{tcolorbox}[colback=blue!10,colframe=blue!50,title=Example: M-bias in Sleep Research,breakable]
\textbf{Study Design}: Recruited volunteers for sleep-brain study from university campus and nearby community\\
\textbf{Selection Process}: Participants self-select based on availability and interest in health research\\
\textbf{Unmeasured Variables}:
\begin{itemize}
    \item Work Schedule: Shift workers have poor sleep \cite{Kecklund2016} but low participation rates in daytime research \cite{Biegus2022}
    \item Education Level: Higher education linked to better white matter integrity \cite{Stern2012,Barulli2013} and higher research participation \cite{Fry2017}
\end{itemize}
\textbf{Result}: Sample over-represents healthy sleepers and under-represents unhealthy sleepers for different reasons\\
\textbf{Bias}: Estimated sleep effects conflate true causal effects with selection-induced associations\\
\textbf{Key Insight}: Controlling for Age does not eliminate bias from unmeasured selection factors
\end{tcolorbox}

\subsection{Types of selection bias}

\citet{Mathur2024Jun} provide a framework for understanding different types of selection bias based on causal graph structures. Their analysis reveals that selection bias can threaten both internal and external validity in distinct ways:

When the exposure (sleep quality) does not affect selection, as in our M-bias sleep example, we can apply simpler definitions of selection bias. For more complex scenarios where exposure affects selection, see \citet{Mathur2024Jun} for detailed frameworks. An {internal bias} occurs when estimates in the selected sample do not reflect true causal effects within that sample, threatening internal generalizability \cite{Mathur2024Jun}. In our M-bias sleep example, internal bias means the estimated sleep effect may not represent the true causal effect even for the specific individuals we studied, because unmeasured variables (Work Schedule and Education Level) create artificial associations through the selection process.

A {net-external bias} occurs when causal effects in the selected sample do not match those in other populations, threatening external generalizability \cite{Mathur2024Jun}. For instance, if our volunteer sample over-represents highly educated individuals (who may respond differently to sleep interventions), then effect estimates may not generalize to the broader population with varying education levels.

\citet{Mathur2024Jun} demonstrate that different graph structures create different combinations of internal and net-external bias (see their Table 2). Researchers can use these graphical criteria to assess which types of bias threaten their specific study design, what assumptions would be required for valid inference, and what conditioning sets could be hypothetically used post-hoc to mitigate these biases.

\subsection{Why drawing DAGs matters for selection bias}

The M-bias example illustrates why explicitly drawing causal graphs is crucial for assessing selection bias. Four key insights emerge:

\paragraph{Standard confounding control may be insufficient.} Even when researchers carefully control for measured common causes like age or biological sex, unmeasured factors influencing selection can still bias results. M-bias specifically shows how controlling for measured common causes of the exposure and outcome (i.e., adjusting for confounders) may not eliminate bias due to unmeasured variables affecting both the outcome and sample selection.

\paragraph{Selection processes are rarely random.} Most neuroimaging studies recruit from convenience samples: university students, hospital patients, or community volunteers. Each recruitment strategy creates different selection pressures that may interact with the exposures and outcomes of interest. Making these selection mechanisms explicit through DAGs helps identify potential bias sources.

\paragraph{The scope of valid inference depends on selection structure.} Different selection mechanisms limit generalizability in different ways. Understanding whether a study faces internal bias, net-external bias, or both determines what conclusions can be drawn and for which populations they apply.

\paragraph{Catch potential selection biases early.} While accounting for selection biases post-hoc is not ideal, it is possible in some cases as long as the the required covariates have been measured. In the example of Figure \ref{fig:collider_examples}(C), measuring education during data collection could suffice as a conditioning variable to eliminate the selection bias. However, after data collection, it may be impossible to make these post-hoc adjustments if these essential conditioning variables are not measured. Therefore, conceptualizing what potential selection biases may exist before data collection can guide choices regarding covariates that may be needed to satisfy identification criteria.

\begin{tcolorbox}[colback=red!10,colframe=red!50,title=Practical Steps for Assessing Selection Bias,breakable]
\textbf{Before Analysis}:\\
1. Draw DAG including recruitment/selection variables\\
2. Think about potential unmeasured factors affecting both exposure/outcome and sample selection\\
3. Potential for selection biases can be identified, and potential control measures assessed, using graphical rules such as \citet{Mathur2024Jun}\\
4. Consider what assumptions (or limitations) are required for your target inference\\
\textbf{During Analysis}:\\
1. Compare sample characteristics to target population where possible\\
2. Test for interactions between exposure effects and selection-related variables\\
3. Consider sensitivity analyses for unmeasured selection factors\\
\textbf{In Interpretation}:\\
1. Clearly specify the population to which conclusions apply\\
2. Acknowledge limitations from selection bias\\
3. Distinguish internal validity (within sample) from external validity (to population)
\end{tcolorbox}

\subsection{Selection bias in common neuroscience research scenarios}

Selection bias manifests across many research research contexts. Understanding these common patterns helps researchers anticipate and address potential bias sources:

\paragraph{Clinical versus healthy control comparisons.} Studies comparing patient populations to healthy controls face inherent selection challenges \cite{Doris2021,Wachinger2020}. Patients are selected based on having the outcome of interest (e.g., depression, ADHD), while controls are selected for lacking clinical symptoms. If brain measures influence both symptom expression and willingness to participate in research, this creates collider bias similar to the professional football example.

\paragraph{Quality control exclusions.} Neuroimaging studies routinely exclude data based on quality metrics like head motion, signal dropout, or processing failures \cite{Power2012,Alexander2016}. When these quality issues are associated with both brain characteristics and behavioral outcomes, exclusion creates selection bias. Excluding high-motion participants may systematically remove individuals with certain behavioral traits, with numerous studies showing that motion exclusion disproportionately affects clinical populations and creates biased samples \cite{Nebel2022Aug,Peverill2025,Cosgrove2022Aug}.

\paragraph{Longitudinal retention.} Studies with multiple timepoints can face differential attrition where dropout is associated with baseline characteristics \cite{Chatfield2005,Glymour2012}. If brain measures influence both the outcome and likelihood of continued participation, analyzing only completers introduces collider bias. This is particularly problematic in aging studies where cognitive decline affects both brain structure and study retention.

\paragraph{Geographic and demographic constraints.} Multi-site studies may recruit different populations at different locations due to geographic, economic, or cultural factors \cite{Wachinger2020}. If brain function or structure influences both outcomes and residential patterns that determine site accessibility, this creates selection bias that standard site adjustment cannot eliminate \cite{Bridgeford2024Feb}.

\subsection{Addressing selection bias: Prevention beats correction}
Unlike confounding bias which, at least in principle, can be eliminated through appropriate adjustment, selection bias can involve fundamental trade-offs \cite{Mathur2024Jun}. When colliders coincide with data quality issues, researchers face a difficult choice: include problematic data (risking measurement error) or exclude it (risking selection bias). Neither option eliminates all bias.

The most effective approaches to selection bias involve prevention through study design rather than post-hoc analytical corrections \cite{Tripepi2010}. Consider these design strategies:

\textbf{Recruitment modifications} can reduce selection pressure. Rather than relying solely on volunteer samples, researchers can use registry-based sampling, partner with community organizations, or implement randomized recruitment strategies across different populations \cite{Charpentier2021}. For sleep studies, this might mean recruiting from both university and community health settings to capture broader socioeconomic representation.

\textbf{Retention strategies} matter for longitudinal studies. Differential dropout creates selection bias when participants who leave the study differ systematically from those who remain \cite{Chatfield2005}. Flexible scheduling, transportation assistance, and maintaining participant engagement can reduce attrition-related bias \cite{Teague2018}.

\textbf{Targeted design elements} might provide reference points. A subset of participants recruited specifically for demographic balance, or measured across multiple sites, could help estimate the magnitude of selection effects and inform correction strategies.

\subsection{Why explicit assumptions matter}

Selection bias often goes unrecognized in standard neuroimaging analyses because recruitment and exclusion processes are treated as methodological details rather than potential bias sources \cite{Charpentier2021}. A neuroimaging study might find associations between sleep quality and white matter integrity in university volunteers, then interpret this as evidence that sleep interventions would benefit the general population. However, recognizing that educated individuals are overrepresented in volunteer samples suggests the findings might not apply to people with different educational backgrounds or life circumstances \cite{Ganguli1998,Charpentier2021}.

These problems do not necessarily invalidate research; they clarify when conclusions may apply to a given target or sample population. Understanding the scope and limitations of findings can be as scientifically valuable as the findings themselves \cite{Rothman2012Dec}. The causal graph framework makes selection processes explicit and allows researchers to evaluate their implications systematically \cite{Mathur2024Jun}.

Drawing causal graphs that include selection variables helps researchers assess whether their study design supports their inferential goals \cite{Mathur2024Jun,hernan2004structural}. This approach does not necessarily require abandoning existing analytical methods, but rather clarifying the assumptions under which those methods yield valid inferences.

\subsection{Moving beyond the assumption of random sampling}

Traditional statistical inference assumes random sampling from a well-defined population. However, this assumption is rarely met in human neuroscience research, where convenience sampling and quality control exclusions are standard practice \cite{Fry2017,LeWinn2017}. Rather than treating this as an unavoidable limitation, the causal inference framework provides tools for understanding and working with non-random samples.

The key insight is that different types of non-random sampling create different bias patterns. Some selection mechanisms threaten internal validity (conclusions about the study sample itself), while others threaten net-external validity (generalization to other populations), and some threaten both \cite{Mathur2024Jun}. By making selection mechanisms explicit through causal graphs, researchers can:

\begin{itemize}[leftmargin=*]
\item Identify which types of bias are most concerning for their research question
\item Determine what additional assumptions would be required for valid inference
\item Design sensitivity analyses appropriate for their selection structure
\item Clearly communicate the scope and limitations of their conclusions
\end{itemize}

\begin{tcolorbox}[colback=red!10,colframe=red!50,title=Section Takeaways,breakable]
\textbf{$\checkmark$ What works:} Recognizing selection bias helps clarify scope of valid inference; some selection patterns allow valid conclusions with explicit assumptions\\
\textbf{$\times$ What fails:} Ignoring selection processes; assuming findings generalize beyond study sample; treating quality control as bias-free\\
\textbf{$\rightarrow$ What to do:} Draw DAGs including selection variables; classify bias types using established frameworks; design studies to minimize problematic selection; clearly specify target populations\\
\textbf{Key insight:} Selection bias often cannot be eliminated analytically; prevention through study design is more effective than post-hoc correction
\end{tcolorbox}

Understanding selection bias represents a crucial complement to confounding control in observational research. While confounding bias can often be addressed through appropriate adjustment strategies, selection bias frequently requires acknowledging fundamental limitations in what can be concluded from a given study design.

%% file: Content/discussion.tex
\section*{Discussion}
\label{sec:disc}

This review demonstrates how principles of causal inference can strengthen neuroscience research by making explicit the assumptions underlying standard statistical analyses. Rather than replacing existing methods, this framework helps researchers understand when multiple regression with covariates supports valid causal interpretations and when additional considerations are needed. Further, we clarify how causal techniques, such as DAGs and positivity-aware methods, can be used to increase transparency of violations of key causal assumptions.

\paragraph{Implications for neuroscience practice}

The scenarios we examined regarding confounding and selection biases represent common challenges in observational neuroscience studies. Each illustrates how seemingly reasonable analytical choices can lead to biased conclusions when identification assumptions are violated.

Confounding control through variable adjustments works well when all common causes of the exposure and outcome are measured and covariate distributions overlap across exposure groups. Our sleep quality example showed how age creates backdoor paths that must be blocked for valid causal inference. A key insight is that standard regression can mask violations of the positivity assumption by extrapolating to covariate combinations that do not exist in the data. Researchers can address this by checking covariate overlap, using propensity-aware methods when appropriate, and conducting sensitivity analyses for unmeasured confounding.

Selection bias often cannot be eliminated through analytical adjustments alone. The M-bias structure in our sleep example illustrated how unmeasured factors affecting both study participation and outcomes can create spurious associations even when measured common causes are controlled. Prevention through study design, such as broader recruitment strategies, balanced inclusion criteria, and retention protocols, is more effective than post-hoc corrections. When selection bias is unavoidable, researchers should clearly specify the population(s) to which their conclusions apply.

Identification trade-offs emerge when multiple bias sources coincide. Quality control procedures that exclude problematic data may trade measurement errors for selection biases, with no perfect solution. Understanding these trade-offs helps researchers make informed decisions about which biases are more acceptable for their specific research questions, or aid in communication of the limitations of a given analysis.

\paragraph{Relationship with existing frameworks}

This review complements recent work on causation across psychology and neuroscience. \citet{Ross2024Feb} provide an important distinction between mechanistic and statistical causation, clarifying that different research questions require different causal approaches. Their framework helps researchers understand when to focus on identifying specific biological mechanisms versus estimating population-level causal effects. \citet{Reid2019Nov} offer a comprehensive review of specialized causal methods for connectivity analysis, including approaches like Granger causality and dynamic causal modeling that can identify directed relationships in time-series neuroimaging data.

Our contribution builds on \citet{rohrer2018thinking}, who introduced DAGs to psychology as tools for making causal assumptions explicit and showed how standard practices like controlling for colliders or mediators can bias results. The present work builds on this framework, specifically focusing on the problem of causal identification, the design implications across the data collection and analysis process, and the statistical implications of failures to do so. Further, we extend this framework to neuroscientific contexts, demonstrating how causal identification principles apply to observational brain-behavior studies.

This framework also connects to recent work highlighting how populations differ in unexpected ways \cite{osayande2025diversity,kopal2023end}. These works demonstrate that population diversity extends beyond typical demographic variables to include factors like handedness interacting with culture, hormone metabolism, and body composition. Our M-bias analysis provides a causal framework for understanding how such unexpected sources of population differences can create bias. When unmeasured (or measured, but unexplored) factors like work schedules or health consciousness influence both brain-behavior relationships and study participation, standard demographic adjustments fail to eliminate bias, helping explain why findings may not generalize even when studies control for obvious common cause variables.

\subsection*{Limitations and future directions}

Approaching neuroscientific questions from a causal perspective cannot solve all possible inferential challenges. Strong identification assumptions are often required, and some research questions may not be amenable to causal inference from observational data alone. However, making these assumptions explicit allows for more informed interpretation of results. 

The examples described herein are heavily simplified for illustrative purposes. In Figure \ref{fig:collider_examples}, for instance, we propose that measuring education level would suffice for post-hoc adjustments that would permit identification of internally and net-externally valid causal estimands. The variables considered for adjustment throughout this review being declared ``sufficient'' to permit causal identification are within the context of the DAGs proposed, and are not necessarily applicable (nor sufficient) for other causal structures.  In addition, we assumed that covariates were measured without appreciable measurement error, which is unlikely to be the case in many studies.

Future work should focus on developing practical tools that make identification checking routine in neuroimaging workflows. This includes software for automated DAG analysis \cite{Textor2016}, improved sensitivity analysis methods for neuroimaging contexts, and study design recommendations that prevent common bias sources.

Our supplement provides more detailed explications for a number of the issues discussed in this review. Appendix \ref{sec:pam_expanded} provides an expanded discussion and comparison of the trade-offs of various methods for adjustment of common cause variables. Appendix \ref{sec:measurement} contextualizes measurement error, a phenomena which loosely occurs when observations of phenomena feature various degrees of systematic or non-systematic bias due to the measurement process. Appendices \ref{sec:batch_effect} and \ref{sec:meas_err}-2 provide hands-on working examples using the ABCD study \cite{Karcher2021Jan} for how to explore datasets which may feature confounding or selection biases, for the problems of multi-site harmonization and head-motion exclusion. Appendix \ref{sec:further_reading} targets several key references, which we believe may be of particular future reading interest to audiences. 

%% file: Content/other_stuff.tex
\paragraph*{Acknowledgments}

This research was supported by the Noyce Foundation through the Women's Brain Health Initiative Data Coordinating Center. During the peer review process, the authors used Claude Sonnet 4.0 (Anthropic \cite{anthropic2024claude4}) as an interactive brainstorming tool to refine the authors' actionable revision strategies, improve manuscript structure, and ensure reviewer feedback was comprehensively addressed. The authors retain full responsibility for the accuracy of the content and recommendations presented herein.

%% file: Content/appendix.tex
\input{Content/appendices/confounding_extended}
\input{Content/appendices/meas_err}

\section{Further reading}
\label{sec:further_reading}
This review provides a brief introduction to causal identification, and contextualizes its importance for neuroscience analyses. For readers wishing for more details, we believe the following textbooks and papers serve as excellent foundations. 

\begin{enumerate}[leftmargin=*]
    \item \textbf{Pearl, J. (2009). \textit{Causality: Models, Reasoning, and Inference} (2nd ed.). Cambridge University Press} \cite{Pearl2009Sep}. The foundational text that established directed acyclic graphs (DAGs) and the backdoor criterion as standard tools for causal reasoning. Provides the theoretical framework underlying all graphical approaches to causal identification.
    
    \item \textbf{Hernán, M. A., \& Robins, J. M. (2020). \textit{Causal Inference: What If}. Chapman \& Hall/CRC} \cite{hernan2020causal}. The standard textbook for applied causal inference, providing accessible explanations of identification assumptions and methods for observational studies. Essential for understanding how causal principles apply to real-world data analysis.
    
    \item \textbf{Stuart, E. A. (2010). Matching methods for causal inference: A review and a look forward. \textit{Statistical Science}, \textit{25}(1), 1--21} \cite{Stuart2010Feb}. Comprehensive review of matching methods and positivity-aware approaches for causal inference. Provides the methodological foundation for alternatives to standard regression when covariate overlap is poor.
    
    \item \textbf{Rohrer, J. M. (2018). Thinking clearly about correlations and causation: Graphical causal models for observational data. \textit{Advances in Methods and Practices in Psychological Science}, \textit{1}(1), 27--42} \cite{rohrer2018thinking}. Introduced DAGs and causal reasoning to psychology, demonstrating how standard practices like controlling for colliders can bias results. Provides the bridge between formal causal inference and psychological research that neuroimaging builds upon.
    
    \item \textbf{Mathur, M. B., \& Shpitser, I. (2025). Simple graphical rules for assessing selection bias in general-population and selected-sample treatment effects. \textit{American Journal of Epidemiology}, \textit{194}(1), 267--277} \cite{Mathur2024Jun}. Provides simple graphical criteria for determining when selection bias threatens causal identification and when covariate adjustment can eliminate bias. Essential for understanding M-bias and other selection mechanisms in neuroimaging studies.
\end{enumerate}

\section{ABCD Data}

Appendix A.4 investigated the reasonableness of positivity assumptions across the sites in the ABCD study. Data were amalgamated using the files and corresponding key in \texttt{participants\_v1.0.3.zip}, retrieved from \href{nda.nih.gov/edit\_collection.html?id=3165}{https://nda.nih.gov/edit\_collection.html?id=3165} \cite{Feczko2021Jul}.

Appendix B.4 groups the demographic covariate information across groups based on the level of average motion from volume-to-volume in the resting state fMRI scan, using the real-time motion tracking information for the resting state fMRI sessions. This was performed by combining the preceding demographic information with the real-time motion tracking information for the resting state fMRI sessions  found in the file \texttt{abcd-data-release-5.0.zip}, in the expanded file \texttt{core/imaging/mri\_y\_qc\_motion.csv}, from the homepage for the ABCD study (non-imaging raw data, June 2023, from \href{nda.nih.gov}{https://nda.nih.gov}). The relevant columns corresponding to the real-time motion tracking information for the resting state fMRI sessions are \texttt{rsfmri\_meanmotion}, ``Resting state fMRI - Average framewise displacement in mm'', and \texttt{rsfmri\_subthreshnvols}, ``Resting state fMRI - Number of frames with FD < 0.2''. We also annotate the raw scores for the CBCL found in the file \texttt{abcd-data-release-5.0.zip}, in the expanded file \texttt{core/mental\_health/mh\_p\_cbcl.csv}. The relevant columns corresponding to the raw and T scores for ADHD were \texttt{cbcl\_scr\_dsm5\_adhd\_r} and \texttt{cbcl\_scr\_dsm5\_adhd\_t}, corresponding to ``Recommended ADHD CBCL DSM5 Scale (raw score)'' and ``Recommended ADHD CBCL DSM5 Scale ($t$ score)'' respectively. The keys for the motion and CBCL data to identify the appropriate columns were retrieved by filtering through \href{data-dict.abcdstudy.org/}{https://data-dict.abcdstudy.org/}. Data are grouped based on whether the mean FD across all volumes exceeds (exclusion group) or is less than (inclusion group) $0.5$.

\paragraph*{Code and Data Availability Statement}

Code and instructions for reproducing the figures and hands-on exploratory analyses contained within this manuscript can be found at \href{https://github.com/ebridge2/causal_neuro}{github.com/ebridge2/causal\_neuro}. Real data exploratory analyses were conducted using data from the ABCD study \href{https://acbdstudy.org}{abcdstudy.org}, a longitudinal multi-site study designed to recruit and follow over $11$,$000$ children into early adulthood. The data obtained for this manuscript were obtained via the NIH Neuroimaging Data Archive (NDA), a permissioned-access repository containing the ABCD study data. Access to this data requires an approved Data use Certification (DUC), which can be obtained via NDA at \href{https://nda.nih.gov/abcd}{nda.nih.gov/abcd}.

\paragraph*{Inclusion and ethics}

This study analyzed existing data from the ABCD study \cite{Karcher2021Jan} accessed through the NIH Neuroimaging Data Archive under an approved Data Use Certification. The original data collection received appropriate IRB approvals with informed consent/assent from all participants and their guardians. Our exploratory analyses indicate that participants from traditionally underserved populations (Black participants, those with lower income and parental education) are disproportionately affected by common quality control procedures in neuroimaging, particularly motion-based exclusion criteria, supporting the results found by previous investigations (e.g., \cite{Cosgrove2022Aug}). Our methodological recommendations aim to enhance inclusivity by identifying approaches that minimize systematic exclusion of underrepresented groups while maintaining scientific rigor. We emphasize that these issues represent both methodological and ethical imperatives, as biased participant selection directly impacts the equitable distribution of benefits from neuroscience research and may produce results that inadequately represent populations with behavioral phenotypes of interest such as ADHD symptoms. The causal inference frameworks we propose aim to make explicit the assumptions underlying neuroimaging analyses, thereby enhancing transparency and scientific integrity.

%% file: Content/appendices/confounding_extended.tex
\section{Controlling for confounding biases (Expanded Discussion)}

\subsection{Methods for estimating causal effects}
\label{sec:pam_expanded}

The most basic approach to analyzing observational data for causal inference is multivariate analysis, typically through multiple regression models \cite{Agresti2015Feb}. An \textit{outcome model} specifies how the outcome variable is influenced by the exposure (the treatment or intervention of interest) and covariates (other variables that might affect the outcome). For instance, in our sleep example, we would model white matter integrity as a function of sleep quality and the covariates (e.g., age and sex via a linear regression). While these models can identify causal estimands when conditional ignorability and positivity assumptions are met, they make violations of key identification assumptions less apparent. As noted by \cite{Antonakis2014May}, if these violations are ignored, resulting analyses can have limited utility for understanding phenomena. Further, consistency of estimators of effects derived from multivariate methods rely heavily on the correctness of the specification of the outcome model, which is often unknown at the time of analysis.

To address some of these limitations, researchers often attempt to employ various techniques which can make more transparent certain types of violations of ID assumptions (typically, positivity). We will broadly refer to this class of methods as \textbf{positivity-aware methods} \footnote{These methods are often characterized as ``causal methods'' or ``causal analyses''; it is important to clarify that positivity-aware methods still require additional ID assumptions to derive causal conclusions, and non-positivity-aware methods can still yield causal inferences. In particular, no method can directly verify from the data the conditional ignorability criterion, and therefore all methods require assumptions and domain expertise regarding the adequacy of observed covariates for causal conclusions.}. The most fundamental is \textit{stratification}, where samples are divided into covariate bins (strata) and analyses are performed within each stratum \cite{Jepsen2004Aug, Rothman2012Dec}. For instance, in our sleep example, prior to performing our multivariate regression, we may first group individuals into sex-matched bins for a specified age range (e.g., one bin might be females between $40$ and $60$, and we might perform regressions of white matter integrity onto sleep quality and age for a given biological sex/age bin). This approach can offer enhanced robustness and make assumption violations transparent; for example, we may have evidence of a positivity violation if a given bin contains only individuals from one sleep group and not the other. However, it can face challenges with dimensionality issues if the covariates are complicated and sensitivity to bin selection \cite{Gelman2006Dec}.

\textbf{Re-weighted methods} offer another approach by changing the strength of each observation's contribution to the analysis, thus making observational studies more like randomized experiments \cite{Stuart2010Feb}. These methods rely on the \textit{propensity score} -- the probability of receiving a particular exposure given the covariates. The \textit{propensity score model}, typically fit using logistic regression or multinomial models, specifies how the exposure depends on covariates, and a fit propensity score model allows us to estimate these propensity scores. Key techniques include \textit{propensity trimming} (PT), which removes samples with extreme propensity scores, and \textbf{inverse probability weighting} (IPW), which weights samples inversely to their propensity scores \cite{Robins1986Jan}. For instance, through inverse probability weighting, we would first estimate propensity scores using a logistic regression of sleep quality onto age and sex, and then incorporate transformations of these propensity scores into a regression of white matter integrity onto sleep quality. Positivity violations are made transparent by these methods, as the propensity scores will be extremely high (or low) for certain samples when the exposure groups differ substantially. More sophisticated approaches include \textit{doubly robust methods} like augmented IPW (AIPW) \cite{bang2005doubly} or targeted maximum likelihood estimation (TMLE) \cite{van2006targeted}, which can provide consistent estimates (estimates that converge to the true value as sample size increases) if either the propensity score model or outcome model is correctly specified. These methods would incorporate the propensity score weights into approaches similar to multivariate methods.

\textbf{Matching methods} take a more direct approach by pairing samples from different exposure groups based on covariate similarity \cite{Stuart2008}. For instance, matching methods might identify participants with good sleep quality, and then for each individual with good sleep, identify individuals with poor sleep quality who have similar covariates to that individual, prior to a regression of white matter integrity onto sleep quality. These methods aim to achieve covariate balance, a condition where the joint covariate distributions are approximately equal across exposure groups \cite{Stuart2010Feb}. In these cases, positivity violations are made transparent through failures to identify matches across the exposure groups due to dissimilarities in the covariates of individual samples. While matching strategies can effectively balance covariate distributions \cite{Ho2011Jun}, they face challenges with high-dimensional data and require careful selection of \textbf{distance metrics} (mathematical measures of covariate similarity between samples).

A crucial limitation across all these methods is their reliance on measuring relevant common causes. The challenge of unobserved confounding (when unmeasured variables affect both exposure and outcome) remains significant, as these methods can only account for measured common causes. Multiple regression is straightforward but can obscure assumption violations regarding positivity, in contrast to positivity-aware methods. Stratification and matching are intuitive, but can be challenging with high-dimensional covariates, and propensity score methods can effectively balance groups but are sensitive to propensity model specification. The choice of method often depends on the specific context and data structure at hand. The assumptions and limitations of different methods are summarized in Table \ref{tab:method_cmp}.

\begin{table}[h]
    \centering
    \includegraphics[width=\linewidth]{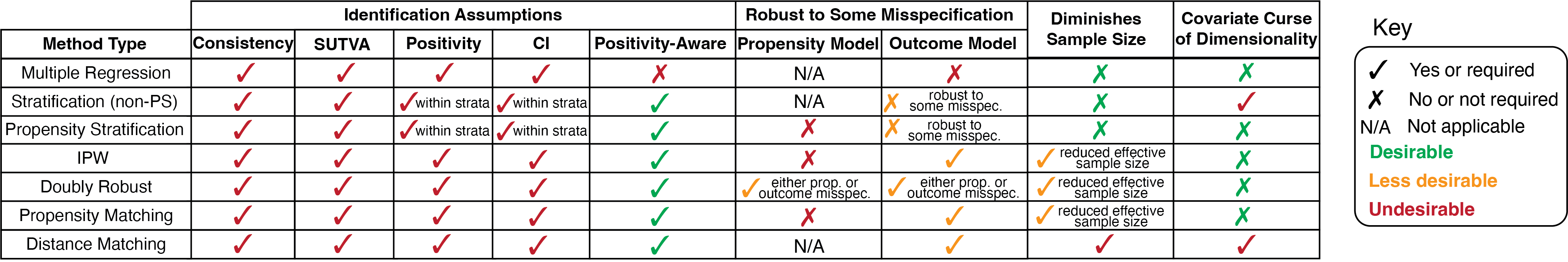}
    \caption{{A comparison of methods for estimating causal effects from observed data, across various dimensions, including the identification assumptions, robustness to certain forms of model misspecification, whether the method reduces the sample size, and whether the method experiences difficulty when the number of covariates or common causes for the adjustment set grows. Note that this chart is for conceptual purposes, and does not reflect every dimension of nuance for each technique.}}
    \label{tab:method_cmp}
\end{table}

\subsection{Advanced considerations}

Standard confounding control assumes that common cause effects are additive and that the exposure effect is constant across all covariate levels. These assumptions may be violated in neuroimaging studies where biological relationships are often complex and interactive.

\paragraph{Effect modification} occurs when the causal effect of sleep on white matter varies across different subgroups. For example, sleep effects might be stronger in older adults whose brains are more vulnerable to vascular challenges, or weaker in younger adults with greater neural plasticity. Traditional regression with additive covariate effects would miss such interactions \cite{Rothman2012Dec}.

Stratified analysis naturally captures effect modification by estimating separate effects within each stratum. Regression approaches can incorporate interactions, but this requires specifying which interactions to include, a decision that should be guided by biological hypotheses rather than data-driven model selection \cite{Gelman2006Dec}.

\paragraph{Nonlinear relationships} between covariates and outcomes violate standard regression assumptions. Age effects on white matter integrity likely follow nonlinear trajectories, with accelerating decline in older adults. Similarly, dose-response relationships between sleep duration and brain health might be U-shaped, with both very short and very long sleep associated with worse outcomes.

Flexible modeling approaches like splines, generalized additive models, or machine learning methods can capture nonlinearities. However, increased model flexibility comes at the cost of reduced interpretability and potential overfitting. The choice between flexible and parametric approaches should balance biological plausibility with statistical power and interpretability. It is important to clarify again that causal relationships can be identified, but still inconsistently estimated, due to difficulties experienced with model selection.

\subsection{Batch effects in multi-site neuroimaging studies}
\label{sec:batch_effect}

Most neuroimaging research relies on observational data where systematic technical differences between measurement sites can confound scientific conclusions. In multi-site consortium studies, researchers pool data from multiple scanning locations to achieve larger sample sizes than possible at single sites. However, this introduces batch effects—systematic technical differences between sites that are unrelated to biological variation of interest \cite{Johnson2007Jan}.

Consider a multi-site study aiming to detect and quantify these batch effects. Even when sites attempt to use identical scanning protocols, systematic differences emerge due to scanner hardware variations, technician practices, local procedures, and site-specific factors. The challenge is determining whether observed differences between sites reflect genuine technical artifacts versus differences in the populations recruited at each site.

\begin{tcolorbox}[colback=blue!10,colframe=blue!50,title=Working Example: Detecting Batch Effects in Multi-site Study,breakable]
\textbf{Research Question}: Do different scanning sites introduce systematic measurement biases in brain connectivity data?\\
\textbf{Study Design}: Cross-sectional analysis of resting-state fMRI from 21 consortium sites\\
\textbf{Exposure}: Scanning site (21 different locations)\\
\textbf{Outcome}: Functional connectivity measurements\\
\textbf{Key confounders}: Demographics (age, biological sex), health status, socioeconomics (education, income)\\
\textbf{Challenge}: Separate technical site effects from demographic differences between site populations\\
\textbf{Real-world example}: ABCD study sites show dramatic income differences: some sites recruit predominantly from high-income families while others recruit from low-income families
\end{tcolorbox}

\subsubsection{Understanding batch effect detection through causal graphs}

\begin{figure}
    \centering
    \includegraphics[width=\linewidth]{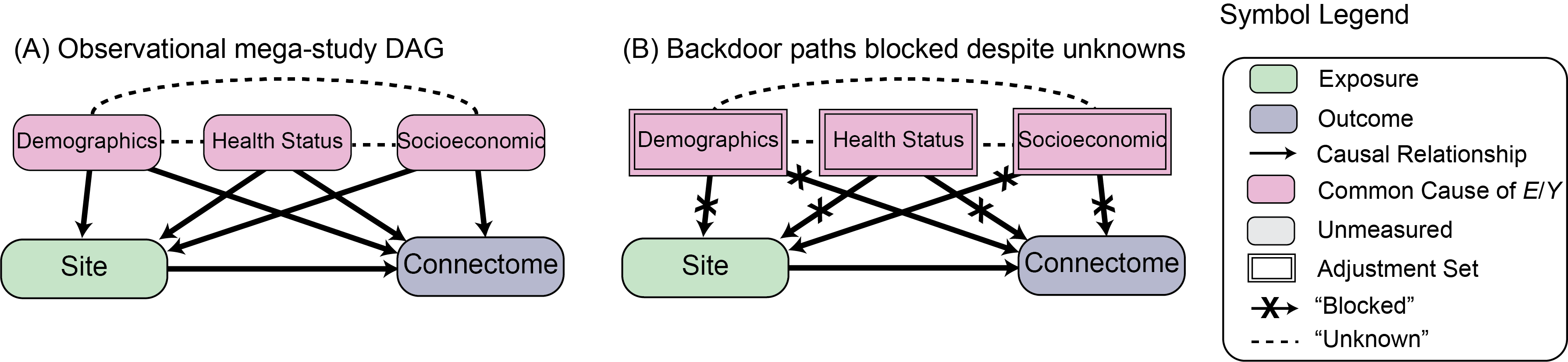}
    \caption{\textbf{Causal structure for batch effect detection}. \textbf{(A)} Site directly affects connectivity measurements (batch effect of interest) but demographics could lead to confounding through differential recruitment, as demographics is a common cause of both the site of measurement and the brain measurements. \textbf{(B)} Controlling for demographics (indicated by the square) blocks the backdoor path, isolating technical site effects.}
    \label{fig:batch_detection}
\end{figure}

Figure \ref{fig:batch_detection}(A) illustrates a potential causal structure for batch effect detection. The exposure of interest is the scanning site itself: we want to estimate the direct causal effect of being scanned at Site A versus Site B on connectivity measurements. This direct effect represents the pure technical batch effect, isolated from any biological differences between participants.

However, demographics creates a backdoor path that confounds this relationship: Site $\leftarrow$ Demographics $\rightarrow$ Connectivity. Different sites tend to recruit participants with different demographic profiles due to their geographic and institutional contexts. Sites located in affluent urban areas may recruit predominantly high-income, highly educated participants, while sites in different regions recruit participants with lower socioeconomic status. Because demographics also directly affect brain connectivity through mechanisms like socioeconomic stress, educational enrichment, and health disparities, unadjusted site comparisons conflate technical artifacts with population differences. 

The goal of batch effect detection is to isolate and test for the causal effect: ``Would connectivity measurements differ if the same participant were scanned at Site A versus Site B?'' This question isolates pure technical differences from potential spillover population differences. However, we will need to make identification assumptions to delineate when an analysis could support these sorts of conclusions.

\subsubsection{Core identification assumptions for batch effect detection}

Estimating causal batch effects requires the four core identification assumptions, but multi-site studies present specific challenges:

\paragraph{Conditional ignorability (No unmeasured confounding)} requires that all variables affecting both site assignment and connectivity measurements are observed and controlled. These variables may, further, have unknown relationships between them (indicated by a dashed line, `--`). For batch effect detection, several key threats are:

\textbf{Demographic differences}: Sites recruit from different populations due to geographic, institutional, and referral patterns. If one site is mostly older individuals and another site is mostly younger indivduals, failing to control for demographics will bias batch effect estimates.

\textbf{Health status differences}: Sites in different regions may have populations with different baseline health profiles due to environmental factors, healthcare access, or lifestyle differences. Health status affects both site assignment and brain measurements, requiring careful measurement and control.

\textbf{Socioeconomic differences}: Site locations correlate with regional socioeconomic patterns. Sites in affluent areas may recruit higher-income, more educated participants than sites in economically disadvantaged regions, and socioeconomic factors could affect brain structure through multiple pathways including stress, nutrition, and cognitive stimulation \cite{Stern2012,Barulli2013}.

\paragraph{Positivity (Overlap)} requires that participants with any given characteristics could plausibly be scanned at multiple sites. Multi-site studies frequently violate this assumption:

If Site A recruits only high-income participants while Site B recruits only low-income participants, there is no socioeconomic overlap between sites. Batch effect estimates would rely entirely on extrapolation, assuming income-connectivity relationships are identical across sites—an untestable assumption.

Even with overlapping income ranges, the joint distribution of characteristics may differ. Site A might recruit high-income, highly educated, healthy participants while Site B recruits lower-income participants with various health and educational backgrounds. Standard regression estimates batch effects by extrapolating to participant types that exist at only one site.

\paragraph{Consistency} requires that site assignment represents a well-defined technical intervention. In multi-site studies, ``being scanned at Site X'' encompasses multiple technical factors:
\begin{itemize}[leftmargin=*]
\item Scanner hardware differences (manufacturer, field strength, gradient performance)
\item Acquisition protocol variations (sequence parameters, resolution, scan duration)  
\item Processing pipeline differences (software versions, parameter settings)
\item Operational differences (positioning procedures, quality control standards)
\end{itemize}

For consistency to hold, researchers should specify whether they are testing for effects across all of these potential factors (e.g., whether there is any causal variability between the sites caused by an amalgam of the aforementioned factors) or specific technical components (e.g., whether there is causal variability specifically due to a particular factor). The choice of question will impact analytic choices regarding sufficient conditioning sets.

\paragraph{No interference} requires careful consideration in multi-site studies. While individual participants' connectivity measurements are typically independent, scanner drift could potentially create interference effects. If scanner hardware performance changes over time due to factors like thermal heating, magnetic field drift, or component aging, participants scanned later in a session or day may have systematically different measurements than those scanned earlier \cite{Friedman2006Jul}. This may violate the no interference assumption because one participant's outcome could depend on the timing and sequence of other participants' scans.

\newpage
\includepdf[pagecommand={\thispagestyle{plain}},  pages=-]{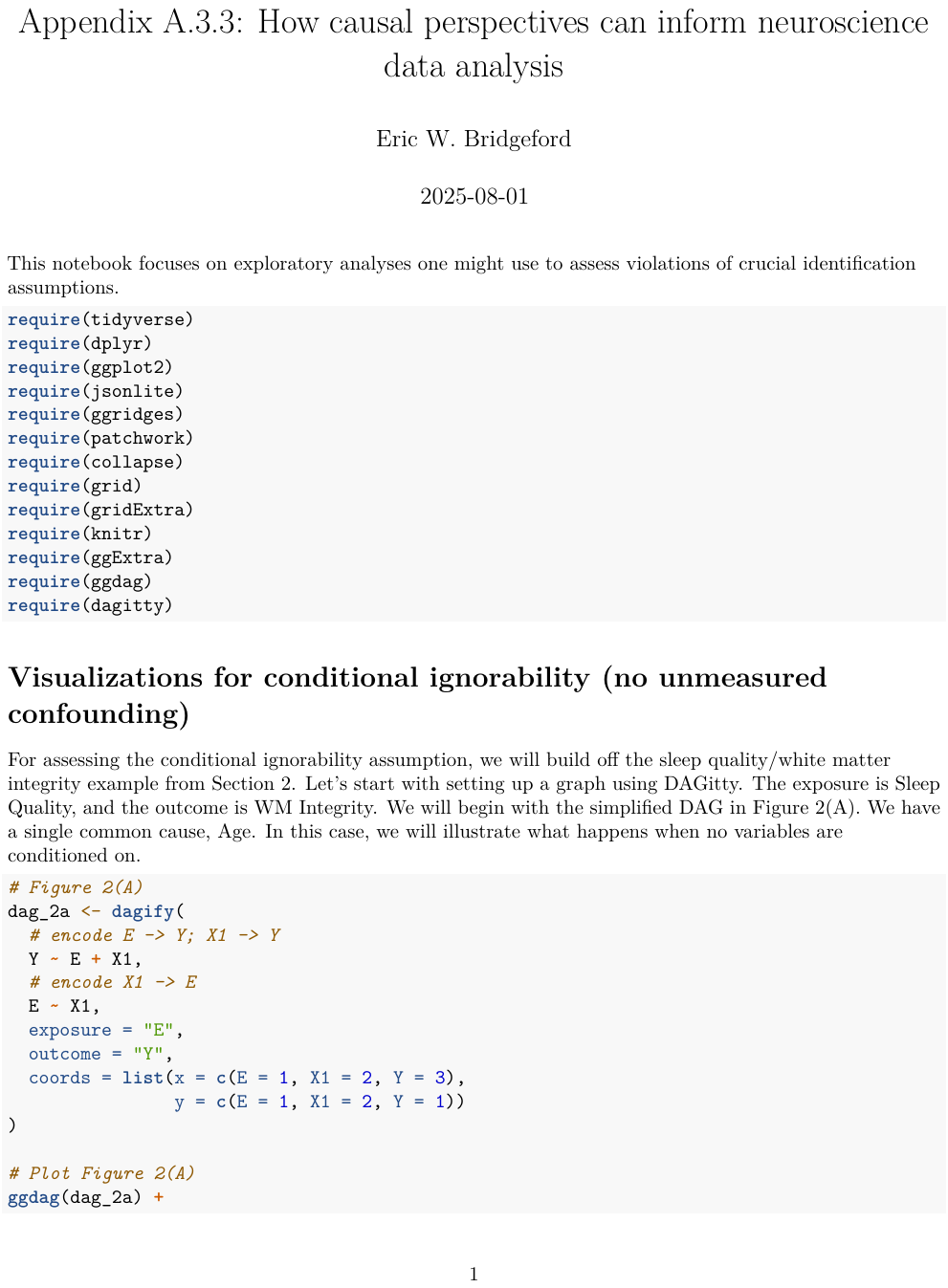}

%% file: Content/appendices/meas_err.tex
\section{Measurement error and motion censoring in neuroimaging}
\label{sec:measurement}

The brain-behavior relationship studies we have examined assume that our measurements accurately capture the underlying biological variables of interest. However, neuroimaging measurements are inevitably corrupted by various sources of noise and artifacts. This section examines how measurement error, particularly from head motion during scanning, creates complex trade-offs between different types of bias that persist even when standard confounding control methods are applied. The key insight is that measurement error becomes a causal inference problem when the measurement process itself has causal structure that correlates systematically with study variables.

\begin{tcolorbox}[colback=gray!10,colframe=gray!50,title=Key Measurement Error Concepts,breakable]
\textbf{Measurement error}: Difference between a measured value and the true underlying value of a phenomenon one wishes to capture\\
\textbf{Non-differential measurement error}: Measurement process satisfies both conditions (i) and (ii) below\\
\textbf{Condition (i)}: The measured exposure is independent of the true outcome conditional on the true exposure\\
\textbf{Condition (ii)}: The measured outcome is independent of the true exposure conditional on the true outcome\\
\textbf{Differential measurement error}: Measurement process violates condition (i) or (ii), creating systematic bias\\
\textbf{Collider stratification bias}: Bias that arises when conditioning on a collider (common outcome of multiple causes)
\end{tcolorbox}

\subsection{Understanding measurement error in causal inference}
\label{sec:meas_err}
Measurement error occurs when there is a difference between a measured value and the true underlying value of a phenomenon one wishes to capture. In neuroimaging studies, we observe connectivity measures ($E^*$) that differ from true neural substrates or functions ($E$) due to head motion, scanner drift, or other artifacts during data acquisition. While all neuroimaging involves measurement error in this sense, the causal inference concern arises when the measurement process systematically correlates with variables in the causal structure under study.

A non-differential measurement error is a measurement error in which, conditional on adjustment covariates \cite{VanderWeele2012Jun}:
\begin{enumerate}[leftmargin=*, label=(\roman*)]
\item the measured exposure is independent of the true outcome conditional on the true exposure, and
\item the measured outcome is independent of the true exposure conditional on the true outcome.
\end{enumerate}

When either condition is violated, we have differential measurement error. Both types of measurement error can create bias in various directions \cite{jurek2005proper,lash2014applying}. However, differential measurement error is particularly concerning for causal inference because the measurement process systematically correlates with the causal variables under study, making it impossible to separate measurement artifacts from the causal relationships of interest \cite{VanderWeele2012Jun}.

\subsubsection{The head motion problem creates differential measurement error}

Head motion during neuroimaging creates differential measurement error because motion is not randomly distributed across participants. Individuals with certain behavioral traits (e.g., Attention Deficit Hyperactive Disorder, ADHD, or Autism Spectrum Disorder, ASD) tend to move more during scanning \cite{Nebel2022Aug,Cosgrove2022Aug}, creating systematic differences in data quality across the groups researchers wish to compare. Movement creates spatial misalignment between successive brain images and introduces spurious signal correlations that cannot be fully corrected by standard preprocessing methods \cite{Power2012,VanDijk2012Jan}.

\begin{tcolorbox}[colback=blue!10,colframe=blue!50,title=Working Example: Motion in Brain-Behavior Studies,breakable]
\textbf{Research Question}: Do neural substrates cause behavioral traits?\\
\textbf{Neural Substrate ($E$)}: Underlying brain structure or connectivity patterns of interest\\
\textbf{Behavioral Traits ($Y$)}: Observable behavioral characteristics that may be influenced by neural substrates\\
\textbf{Head Motion ($R$)}: Participant movement during scanning, influenced by behavioral traits and demographics\\
\textbf{Connectivity Measure ($E^*$)}: Observed brain measurements corrupted by motion artifacts\\
\textbf{Key Challenge}: Motion creates differential measurement error, but conditioning on motion to address this introduces selection bias when combined with necessary demographic adjustments
\end{tcolorbox}

This creates a causal structure where neural substrates influence behavioral traits, and these same behavioral traits affect head motion during scanning. Head motion then corrupts our measurements of the neural substrates, violating condition (i) for non-differential measurement error because our measured connectivity depends on behavioral traits through the motion pathway, independent of the true neural substrate.

\subsection{Motion censoring: Trading measurement error for selection bias}

\begin{figure}
   \centering
   \includegraphics[width=\linewidth]{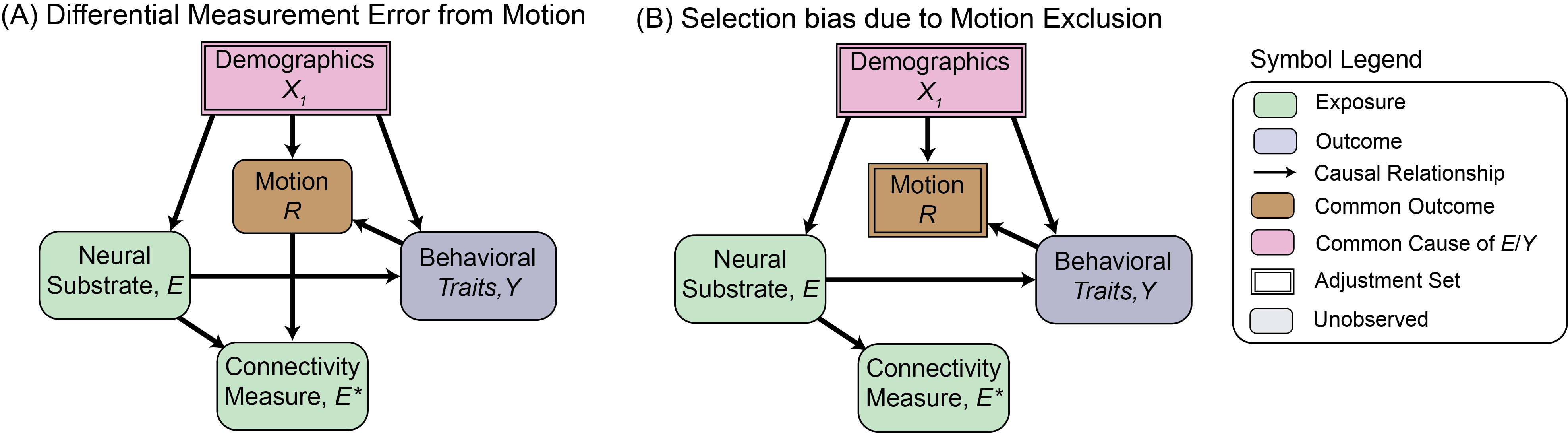}
   \caption{\textbf{The measurement-selection trade-off in brain-behavior studies.} \textbf{(A)} Differential measurement error with demographic conditioning: Demographics must be conditioned on (square) to control for common causes, but head motion varies with behavioral traits, creating systematic measurement differences across groups in the connectivity measure. \textbf{(B)} Motion conditioning trades measurement error for selection bias: Conditioning on head motion (square) eliminates differential measurement error but creates collider stratification bias because head motion is influenced by both conditioned variables (demographics and behavioral traits).}
   \label{fig:motion_measurement}
\end{figure}

Figure \ref{fig:motion_measurement} illustrates this methodological trade-off. Figure \ref{fig:motion_measurement}(A) shows the reality of brain-behavior studies: there is a backdoor path Neural Activation $\leftarrow$ Demographics $\rightarrow$ Behavioral Traits, which could be blocked by conditioning on Demographics (as established in Section \ref{sec:confounding}). However, in this scenario, Motion introduces differential measurement error because data quality varies systematically with behavioral traits. Figure \ref{fig:motion_measurement}(B) shows a potential solution: conditioning on head motion to eliminate the differential measurement error (i.e., after conditioning on low motion individuals, motion artifacts are no longer present in the Connectivity Measures $Y^*$). However, this creates selection bias because head motion becomes a collider when we simultaneously condition on both demographics (necessary for common cause control) and head motion (to address measurement error): we end up disproportionately excluding individuals of certain demographic groups \cite{Cosgrove2022Aug,Nebel2022Aug} and symptom presentations \cite{Nebel2022Aug} compared to our targets. This conditioning strategy can create spurious associations between our surrogate exposure (Connectivity Measurements) and our outcome (Behavioral Traits).

Standard neuroimaging practice involves motion censoring through individual-level exclusion, timepoint scrubbing, or weighted approaches. Each represents a form of conditioning on head motion aimed at improving data quality by addressing differential measurement error. However, when combined with necessary demographic adjustments for common cause control, these approaches create the competing bias structures illustrated in Figure \ref{fig:motion_measurement}.

\subsubsection{An unsolved methodological challenge}

This represents a fundamental dilemma where addressing differential measurement error through certain means (e.g., conditioning on motion) while controlling for common causes (e.g., conditioning on demographics) simultaneously creates a selection bias. This methodological challenge cannot be resolved through standard analytical adjustments alone, as each approach to motion handling introduces different types of bias that may compromise causal conclusions.

The juxtaposition of measurement error and selection bias in motion-corrupted neuroimaging data represents a significant methodological gap that we believe warrants prioritized attention through a causal inference lens. This problem is particularly acute because current standard practices in the field routinely involve motion censoring decisions that create these trade-offs, yet the causal implications are rarely made explicit.

\paragraph{A priority area for causal methodology} The epidemiological literature offers approaches for addressing measurement error, particularly regression calibration methods where estimates are calibrated to account for measurement errors \cite{carroll1995measurement}. While most existing methods address non-differential measurement errors, recent efforts have shown promise in developing techniques for differential measurement errors \cite{leCessie2012Jul}. However, neuroimaging data presents unique challenges due to its high-dimensionality and the simultaneous presence of measurement error and collider structures. We posit that the development of causally-inspired methods specifically designed for neuroimaging contexts represents a critical area for future methodological research.

\paragraph{Current practice requires explicit acknowledgment} Until more robust methodological solutions are developed, studies that implement motion exclusion practices should explicitly acknowledge the potential for both measurement error and selection bias. When motion censoring disproportionately affects groups of interest, this represents a major limitation that should be prominently discussed rather than treated as routine preprocessing. Current reporting practices often treat motion exclusion as purely technical, but from a causal perspective, these decisions fundamentally alter the study population and the causal questions that can be reliably answered.

\paragraph{Design-based solutions merit investigation} Addressing these challenges through study design—such as motion-tolerant sequences (e.g., continuing to improve motion correction techniques), behavioral interventions to reduce motion, shorter scan sessions, or sampling strategies that explicitly balance motion across groups, may be more effective than post-hoc analytical corrections. However, the relative effectiveness of different design approaches for preserving causal interpretability remains an understudied area requiring systematic evaluation.

\begin{tcolorbox}[colback=red!10,colframe=red!50,title=Section Takeaways,breakable]
\textbf{$\checkmark$ What works:} Recognizing that measurement error becomes problematic when systematically related to study variables; epidemiological approaches provide foundation for future methodological development\\
\textbf{$\times$ What fails:} Current motion censoring practices often ignore causal implications; no existing approach can simultaneously eliminate both differential measurement error and selection bias\\  
\textbf{$\rightarrow$ What to do:} Prioritize development of causally-informed methods for neuroimaging; explicitly acknowledge motion exclusion as creating fundamental trade-offs; investigate design-based prevention strategies\\
\textbf{Key insight:} Motion censoring creates unavoidable trade-offs between measurement accuracy and sample representativeness that represent a critical gap requiring methodological innovation
\end{tcolorbox}

Understanding measurement error and its interaction with selection processes represents a crucial consideration for neuroimaging studies. These challenges illustrate why causal inference requires explicit consideration of data generation mechanisms, not just statistical modeling choices, and highlight motion censoring as a priority area for future methodological development.

\newpage
\includepdf[pagecommand={\thispagestyle{plain}},  pages=-]{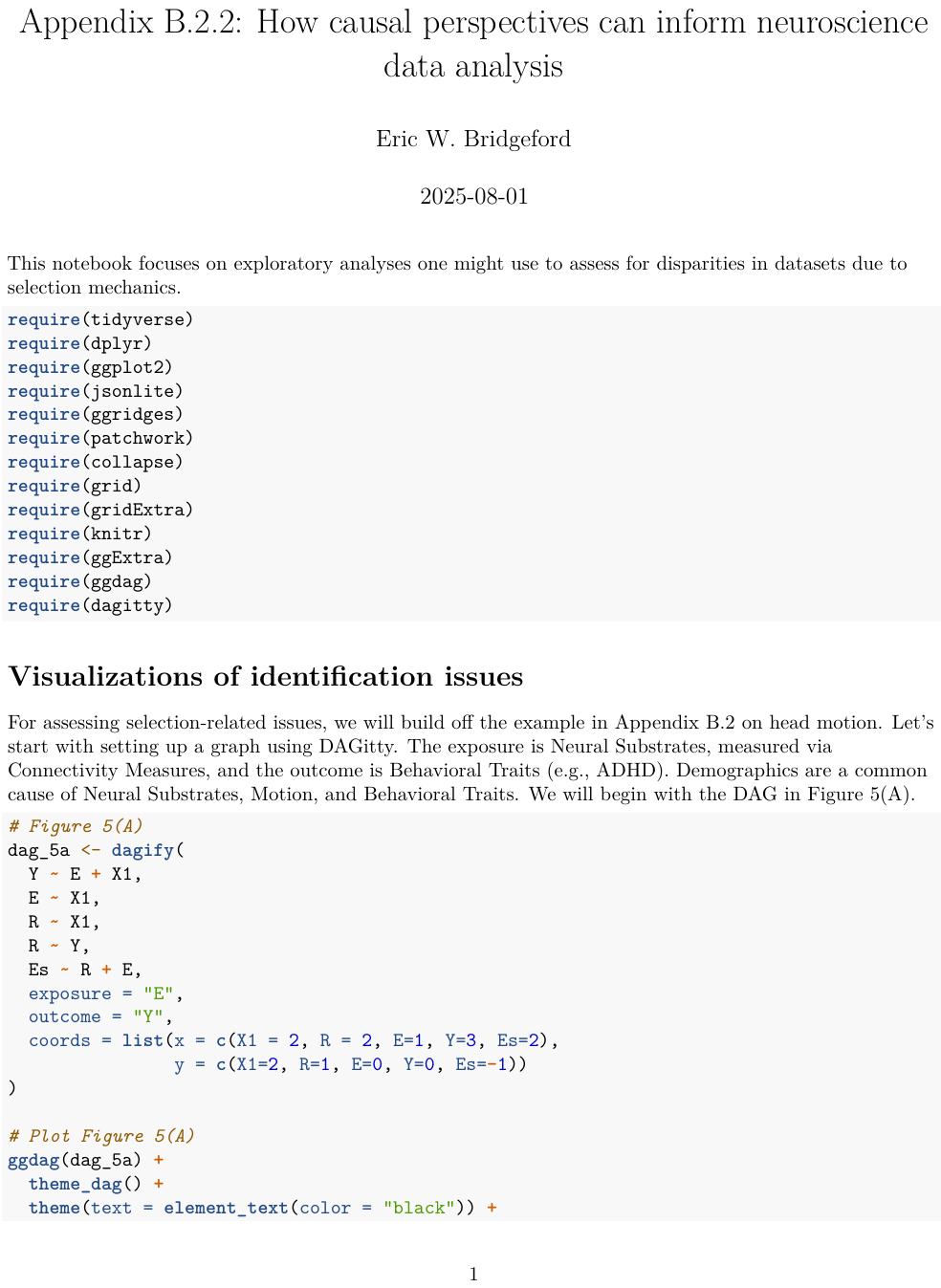}

\subsection{Temporal ordering and reverse causation}

\paragraph{Temporal ordering and reverse causation} Cross-sectional brain-behavior studies face an additional challenge: determining which neural changes are causes versus consequences of disease progression. Many examine associations between neural substrates and behavioral outcomes in cross-sectional designs where the temporal ordering of variables remains unknown. 

Consider a simplified scenario: a healthy individual initially shows normal indicators for two arbitrary neurological biomarkers ($x_1 = 0$, $x_2 = 0$). The true causal pathway might begin with subtle changes in one marker ($x_1 = 1$) leading to diagnosable illness, after which both markers show more dramatic changes ($x_1 = 2$, $x_2 = 1$). In a cross-sectional study observing only disease state, it becomes impossible to determine which connectivity changes caused behavioral traits or were downstream consequences of illness progression, as statistical analyses alone cannot adjudicate between these possibilities.

This temporal ambiguity compounds the measurement error challenges discussed above. Even if motion censoring could perfectly address differential measurement error without creating selection bias, the resulting associations might reflect disease consequences rather than disease causes. While modeling in a theoretically appropriate direction does not guarantee that causal relationships can be identified, researchers can at least delineate assumptions under which identification might be possible.

\paragraph{Measurement error in covariates compounds the problem} The challenges discussed above are further complicated when the adjustment variables themselves contain measurement error. \citet{westfall2016statistically} demonstrated that when covariates used for statistical control have measurement error, standard regression approaches can exhibit extremely high false positive rates. Counterintuitively, they found that ``error rates are highest—in some cases approaching 100\%—when sample sizes are large and reliability is moderate''. This creates a compounding problem in neuroimaging studies: not only do motion artifacts introduce differential measurement error in brain measures, but demographic and behavioral variables commonly used for adjustment (such as socioeconomic status questionnaires, cognitive assessments, or symptom severity scales) typically contain substantial measurement error themselves. When multiple imperfectly measured predictors are included in regression models, ''measurement unreliability and model misspecification will often have a deleterious and large effect on parameter estimates (and associated error rates)''. The interaction between measurement error in both brain measures and covariates creates a methodological challenge that extends well beyond the motion censoring trade-offs discussed above, potentially undermining causal conclusions even when individual measurement issues seem manageable.

%% file: causal_perspective.bbl
\begin{thebibliography}{91}
\providecommand{\natexlab}[1]{#1}
\providecommand{\url}[1]{\texttt{#1}}
\expandafter\ifx\csname urlstyle\endcsname\relax
  \providecommand{\doi}[1]{doi: #1}\else
  \providecommand{\doi}{doi: \begingroup \urlstyle{rm}\Url}\fi

\bibitem[Bechtel and Abrahamsen(2005)]{Bechtel2005Jun}
William Bechtel and Adele Abrahamsen.
\newblock {Explanation: a mechanist alternative}.
\newblock \emph{Studies in History and Philosophy of Science Part C: Studies in History and Philosophy of Biological and Biomedical Sciences}, 36\penalty0 (2):\penalty0 421--441, June 2005.
\newblock \doi{10.1016/j.shpsc.2005.03.010}.

\bibitem[Ross and Bassett(2024)]{Ross2024Feb}
Lauren~N. Ross and Dani~S. Bassett.
\newblock {Causation in neuroscience: keeping mechanism meaningful}.
\newblock \emph{Nature Reviews Neuroscience}, 25:\penalty0 81--90, February 2024.
\newblock \doi{10.1038/s41583-023-00778-7}.

\bibitem[Reid et~al.(2019)Reid, Headley, Mill, Sanchez-Romero, Uddin, Marinazzo, Lurie, Vald{\ifmmode\acute{e}\else\'{e}\fi}s-Sosa, Hanson, Biswal, Calhoun, Poldrack, and Cole]{Reid2019Nov}
Andrew~T. Reid, Drew~B. Headley, Ravi~D. Mill, Ruben Sanchez-Romero, Lucina~Q. Uddin, Daniele Marinazzo, Daniel~J. Lurie, Pedro~A. Vald{\ifmmode\acute{e}\else\'{e}\fi}s-Sosa, Stephen~Jos{\ifmmode\acute{e}\else\'{e}\fi} Hanson, Bharat~B. Biswal, Vince Calhoun, Russell~A. Poldrack, and Michael~W. Cole.
\newblock {Advancing functional connectivity research from association to causation}.
\newblock \emph{Nature Neuroscience}, 22:\penalty0 1751--1760, November 2019.
\newblock \doi{10.1038/s41593-019-0510-4}.

\bibitem[Osayande et~al.(2025)Osayande, Marotta, Aggarwal, Kopal, Holmes, Yip, and Bzdok]{osayande2025diversity}
Nicole Osayande, Justin Marotta, Shambhavi Aggarwal, Jakub Kopal, Avram Holmes, Sarah~W. Yip, and Danilo Bzdok.
\newblock Quantifying associations between socio-spatial factors and cognitive development in the abcd cohort.
\newblock \emph{Nature Computational Science}, 5\penalty0 (3):\penalty0 221--233, March 2025.
\newblock \doi{10.1038/s43588-025-00774-0}.

\bibitem[Kopal et~al.(2023)Kopal, Uddin, and Bzdok]{kopal2023end}
Jakub Kopal, Lucina~Q. Uddin, and Danilo Bzdok.
\newblock The end game: respecting major sources of population diversity.
\newblock \emph{Nature Methods}, 20\penalty0 (8):\penalty0 1122--1128, March 2023.
\newblock \doi{10.1038/s41592-023-01812-3}.

\bibitem[Mander et~al.(2017)Mander, Winer, and Walker]{Mander2017}
Bryce~A. Mander, Joseph~R. Winer, and Matthew~P. Walker.
\newblock Sleep and human aging.
\newblock \emph{Neuron}, 94\penalty0 (1):\penalty0 19--36, April 2017.
\newblock \doi{10.1016/j.neuron.2017.02.004}.

\bibitem[Sexton et~al.(2014)Sexton, Storsve, Walhovd, Johansen-Berg, and Fjell]{Sexton2014}
Claire~E. Sexton, Andreas~B. Storsve, Kristine~B. Walhovd, Heidi Johansen-Berg, and Anders~M. Fjell.
\newblock Poor sleep quality is associated with increased cortical atrophy in community-dwelling adults.
\newblock \emph{Neurology}, 83\penalty0 (11):\penalty0 967--973, 2014.
\newblock \doi{10.1212/wnl.0000000000000774}.

\bibitem[Fjell et~al.(2014)Fjell, McEvoy, Holland, Dale, and Walhovd]{Fjell2014}
Anders~M. Fjell, Linda McEvoy, Dominic Holland, Anders~M. Dale, and Kristine~B. Walhovd.
\newblock What is normal in normal aging? effects of aging, amyloid and alzheimer’s disease on the cerebral cortex and the hippocampus.
\newblock \emph{Progress in Neurobiology}, 117:\penalty0 20--40, 2014.
\newblock \doi{10.1016/j.pneurobio.2014.02.004}.

\bibitem[Xie et~al.(2013)Xie, Kang, Xu, Chen, Liao, Thiyagarajan, O’Donnell, Christensen, Nicholson, Iliff, Takano, Deane, and Nedergaard]{Xie2013}
Lulu Xie, Hongyi Kang, Qiwu Xu, Michael~J. Chen, Yonghong Liao, Meenakshisundaram Thiyagarajan, John O’Donnell, Daniel~J. Christensen, Charles Nicholson, Jeffrey~J. Iliff, Takahiro Takano, Rashid Deane, and Maiken Nedergaard.
\newblock Sleep drives metabolite clearance from the adult brain.
\newblock \emph{Science}, 342\penalty0 (6156):\penalty0 373--377, October 2013.
\newblock \doi{10.1126/science.1241224}.

\bibitem[Hill(1965)]{Hill1965May}
Austin~Bradford Hill.
\newblock The environment and disease: Association or causation?
\newblock \emph{Proceedings of the Royal Society of Medicine}, 58\penalty0 (5):\penalty0 295--300, May 1965.
\newblock \doi{10.1177/003591576505800503}.

\bibitem[Gupta(2011)]{gupta2011intention}
Sandeep~K Gupta.
\newblock Intention-to-treat concept: A review.
\newblock \emph{Perspectives in Clinical Research}, 2\penalty0 (3):\penalty0 109, 2011.
\newblock \doi{10.4103/2229-3485.83221}.

\bibitem[Fisher et~al.(1990)Fisher, Dixon, Herson, Frankowski, Hearron, and Peace]{fisher1990intention}
Lloyd~D Fisher, Dennis~O Dixon, Jay Herson, Ralph~K Frankowski, Marion~S Hearron, and Karl~E Peace.
\newblock \emph{Statistical issues in drug research and development}, chapter Intention to treat in clinical trials, pages 331 -- 350.
\newblock CRC Press, 1990.

\bibitem[Sibbald and Roberts(1998)]{sibbald1998understanding}
B.~Sibbald and C.~Roberts.
\newblock Understanding controlled trials: Crossover trials.
\newblock \emph{BMJ}, 316\penalty0 (7146):\penalty0 1719--1720, 1998.
\newblock \doi{10.1136/bmj.316.7146.1719}.

\bibitem[Rubin(1974)]{rubin1974estimating}
Donald~B. Rubin.
\newblock Estimating causal effects of treatments in randomized and nonrandomized studies.
\newblock \emph{Journal of Educational Psychology}, 66\penalty0 (5):\penalty0 688--701, October 1974.
\newblock \doi{10.1037/h0037350}.

\bibitem[Hern{\'a}n and Robins(2020)]{hernan2020causal}
Miguel~A. Hern{\'a}n and James~M. Robins.
\newblock \emph{Causal Inference: What If}.
\newblock Chapman \& Hall/CRC, Boca Raton, 2020.
\newblock ISBN 9781420076165.

\bibitem[Ciolino et~al.(2013)Ciolino, Martin, Zhao, Jauch, Hill, and Palesch]{Ciolino2013}
Jody~D. Ciolino, Rene{\ifmmode\acute{e}\else\'{e}\fi}~H. Martin, Wenle Zhao, Edward~C. Jauch, Michael~D. Hill, and Yuko~Y. Palesch.
\newblock {Covariate Imbalance and Adjustment for Logistic Regression Analysis of Clinical Trial Data}.
\newblock \emph{Journal of biopharmaceutical statistics}, 23\penalty0 (6):\penalty0 1383, 2013.
\newblock \doi{10.1080/10543406.2013.834912}.

\bibitem[Power et~al.(2017)Power, Parkhill, and de~Oliveira]{Power2017Jan}
Robert~A. Power, Julian Parkhill, and Tulio de~Oliveira.
\newblock {Microbial genome-wide association studies: lessons from human GWAS}.
\newblock \emph{Nature Reviews Genetics}, 18:\penalty0 41--50, January 2017.
\newblock \doi{10.1038/nrg.2016.132}.

\bibitem[Pearl(2009)]{Pearl2009Sep}
Judea Pearl.
\newblock \emph{{Causality: Models, Reasoning and Inference}}.
\newblock Cambridge University Press, Cambridge, England, UK, 2nd edition, September 2009.
\newblock \doi{10.5555/1642718}.

\bibitem[PEARL(1995)]{pearl1995causal}
JUDEA PEARL.
\newblock Causal diagrams for empirical research.
\newblock \emph{Biometrika}, 82\penalty0 (4):\penalty0 669--688, 1995.
\newblock \doi{10.1093/biomet/82.4.669}.

\bibitem[Greenland(1996)]{Greenland1996May}
S.~Greenland.
\newblock {Confounding and exposure trends in case-crossover and case-time-control designs}.
\newblock \emph{Epidemiology (Cambridge, Mass.)}, 7\penalty0 (3):\penalty0 231--239, May 1996.
\newblock \doi{10.1097/00001648-199605000-00003}.

\bibitem[Imbens and Rubin(2015)]{Imbens2015Apr}
Guido~W. Imbens and Donald~B. Rubin.
\newblock \emph{Causal Inference for Statistics, Social, and Biomedical Sciences}.
\newblock Cambridge University Press, April 2015.
\newblock \doi{10.1017/cbo9781139025751}.

\bibitem[Rosenbaum and Rubin(1983)]{rosenbaum1983central}
Paul~R Rosenbaum and Donald~B Rubin.
\newblock The central role of the propensity score in observational studies for causal effects.
\newblock \emph{Biometrika}, 70\penalty0 (1):\penalty0 41--55, 1983.
\newblock \doi{10.1093/biomet/70.1.41}.

\bibitem[Sehgal and Mignot(2011)]{Sehgal2011}
Amita Sehgal and Emmanuel Mignot.
\newblock Genetics of sleep and sleep disorders.
\newblock \emph{Cell}, 146\penalty0 (2):\penalty0 194--207, 2011.

\bibitem[Dashti et~al.(2019)Dashti, Jones, Wood, et~al.]{Dashti2019}
Hassan~S Dashti, Samuel~E Jones, Andrew~R Wood, et~al.
\newblock Genome-wide association study identifies genetic loci for self-reported habitual sleep duration supported by accelerometer-derived estimates.
\newblock \emph{Nature Communications}, 10\penalty0 (1):\penalty0 1100, 2019.

\bibitem[Papadopoulos and Sosso(2023)]{Grandner2022}
Dimitrios Papadopoulos and Faustin Sosso.
\newblock Socioeconomic status and sleep health: a narrative synthesis of 3 decades of empirical research.
\newblock \emph{Journal of Clinical Sleep Medicine}, 2023.

\bibitem[Li et~al.(2023)Li, Cai, Taylor, Eisenstein, Barch, Marek, and Hershey]{Shaked2022}
Zhaolong~Adrian Li, Yuqi Cai, Rita~L. Taylor, Sarah~A. Eisenstein, Deanna~M. Barch, Scott Marek, and Tamara Hershey.
\newblock Associations between socioeconomic status, obesity, cognition, and white matter microstructure in children.
\newblock \emph{JAMA Network Open}, 6\penalty0 (6):\penalty0 e2320276, 2023.
\newblock \doi{10.1001/jamanetworkopen.2023.20276}.

\bibitem[Wang et~al.(2020)Wang, Wang, Wei, Xia, Tian, Cui, and Li]{wang2020relbtwn}
Dan-Qiong Wang, Lei Wang, Miao-Miao Wei, Xiao-Shuang Xia, Xiao-Lin Tian, Xiao-Hong Cui, and Xin Li.
\newblock Relationship between type 2 diabetes and white matter hyperintensity: A systematic review.
\newblock \emph{Frontiers in Endocrinology}, 11, December 2020.
\newblock \doi{10.3389/fendo.2020.595962}.

\bibitem[Hollocks et~al.(2015)Hollocks, Lawrence, Brookes, Barrick, Morris, Husain, and Markus]{Hollocks2015}
Matthew~J. Hollocks, Andrew~J. Lawrence, Rebecca~L. Brookes, Thomas~R. Barrick, Robin~G. Morris, Masud Husain, and Hugh~S. Markus.
\newblock Differential relationships between apathy and depression with white matter microstructural changes and functional outcomes.
\newblock \emph{Brain}, 138\penalty0 (12):\penalty0 3803--3815, October 2015.
\newblock \doi{10.1093/brain/awv304}.

\bibitem[Andrade(2020)]{Andrade2021}
Chittaranjan Andrade.
\newblock The inconvenient truth about convenience and purposive samples.
\newblock \emph{Indian Journal of Psychological Medicine}, 43\penalty0 (1):\penalty0 86--88, December 2020.

\bibitem[LeWinn et~al.(2017)LeWinn, Sheridan, Keyes, Hamilton, and McLaughlin]{LeWinn2017}
Kaja~Z. LeWinn, Margaret~A. Sheridan, Katherine~M. Keyes, Ava Hamilton, and Katie~A. McLaughlin.
\newblock Sample composition alters associations between age and brain structure.
\newblock \emph{Nature Communications}, 8\penalty0 (1), October 2017.

\bibitem[Cole and Frangakis(2009)]{cole2009consistency}
Stephen~R Cole and Constantine~E Frangakis.
\newblock Consistency statement in causal inference: definition, identification and estimation.
\newblock \emph{Epidemiology}, 20\penalty0 (1):\penalty0 3--5, 2009.

\bibitem[Rubin(1980)]{rubin1980randomization}
Donald~B Rubin.
\newblock Randomization analysis of experimental data: The fisher randomization test comment.
\newblock \emph{Journal of the American Statistical Association}, 75\penalty0 (371):\penalty0 591--593, 1980.

\bibitem[Rubin(1986)]{rubin1986comment}
Donald~B Rubin.
\newblock Comment: Which ifs have causal answers.
\newblock \emph{Journal of the American Statistical Association}, 81\penalty0 (396):\penalty0 961--962, 1986.

\bibitem[Stuart(2010)]{Stuart2010Feb}
Elizabeth~A. Stuart.
\newblock Matching methods for causal inference: A review and a look forward.
\newblock \emph{Statistical Science}, 25\penalty0 (1), February 2010.
\newblock \doi{10.1214/09-sts313}.

\bibitem[Agresti(2015)]{Agresti2015Feb}
Alan Agresti.
\newblock \emph{{Foundations of Linear and Generalized Linear Models (Wiley Series in Probability and Statistics)}}.
\newblock Wiley, Hoboken, NJ, USA, February 2015.
\newblock ISBN 978-1-11873003-4.

\bibitem[Antonakis et~al.(2014)Antonakis, Bendahan, Jacquart, and Lalive]{Antonakis2014May}
John Antonakis, Samuel Bendahan, Philippe Jacquart, and Rafael Lalive.
\newblock \emph{The Oxford Handbook of Leadership and Organizations}, chapter {Causality and Endogeneity: Problems and Solutions}.
\newblock Oxford University Press, May 2014.
\newblock \doi{10.1093/oxfordhb/9780199755615.013.007}.

\bibitem[Jepsen et~al.(2004)Jepsen, Johnsen, Gillman, and S{\o}rensen]{Jepsen2004Aug}
P.~Jepsen, S.~P. Johnsen, M.~W. Gillman, and H.~T. S{\o}rensen.
\newblock {Interpretation of observational studies}.
\newblock \emph{Heart}, 90\penalty0 (8):\penalty0 956--960, August 2004.
\newblock \doi{10.1136/hrt.2003.017269}.

\bibitem[Rothman et~al.(2012)Rothman, Greenland, and Lash]{Rothman2012Dec}
Kenneth~J Rothman, Sander Greenland, and Timothy~L Lash.
\newblock \emph{Modern Epidemiology}.
\newblock Lippincott Williams \& Wilkins, Philadelphia, 3rd edition, 2012.

\bibitem[Gelman and Hill(2006)]{Gelman2006Dec}
Andrew Gelman and Jennifer Hill.
\newblock \emph{{Data Analysis Using Regression and Multilevel/Hierarchical Models}}.
\newblock Cambridge University Press, Cambridge, England, UK, December 2006.
\newblock ISBN 978-0521686891.
\newblock \doi{10.1017/CBO9780511790942}.

\bibitem[Robins(1986)]{Robins1986Jan}
James Robins.
\newblock A new approach to causal inference in mortality studies with a sustained exposure period -- application to control of the healthy worker survivor effect.
\newblock \emph{Mathematical Modelling}, 7\penalty0 (9--12):\penalty0 1393--1512, 1986.
\newblock \doi{10.1016/0270-0255(86)90088-6}.

\bibitem[Stuart and Rubin(2008)]{Stuart2008}
Elizabeth~A. Stuart and Donald~B. Rubin.
\newblock \emph{{Best Practices in Quantitative Methods}}, chapter Best practices in quasi-experimental designs.
\newblock SAGE Publications, Inc., Thousand Oaks, CA, USA, 2008.
\newblock ISBN 978-1-41299562-7.
\newblock \doi{10.4135/9781412995627}.

\bibitem[Textor et~al.(2017)Textor, van~der Zander, Gilthorpe, Liśkiewicz, and Ellison]{Textor2016}
Johannes Textor, Benito van~der Zander, Mark~S. Gilthorpe, Maciej Liśkiewicz, and George~T.H. Ellison.
\newblock Robust causal inference using directed acyclic graphs: the r package ‘dagitty’.
\newblock \emph{International Journal of Epidemiology}, page dyw341, January 2017.
\newblock \doi{10.1093/ije/dyw341}.

\bibitem[Greenland(2005)]{Greenland2005Mar}
Sander Greenland.
\newblock Multiple-bias modelling for analysis of observational data.
\newblock \emph{Journal of the Royal Statistical Society Series A: Statistics in Society}, 168\penalty0 (2):\penalty0 267--306, March 2005.
\newblock \doi{10.1111/j.1467-985x.2004.00349.x}.

\bibitem[VanderWeele and Ding(2017)]{VanderWeele2017Aug}
Tyler~J. VanderWeele and Peng Ding.
\newblock {Sensitivity Analysis in Observational Research: Introducing the E-Value}.
\newblock \emph{Annals of Internal Medicine}, 167\penalty0 (4):\penalty0 268--274, August 2017.
\newblock \doi{10.7326/M16-2607}.

\bibitem[Cox et~al.(2015)Cox, MacPherson, Ferguson, Royle, Maniega, Hernández, Bastin, MacLullich, Wardlaw, and Deary]{cox2016white}
Simon~R. Cox, Sarah~E. MacPherson, Karen~J. Ferguson, Natalie~A. Royle, Susana~Muñoz Maniega, Maria del C.~Valdés Hernández, Mark~E. Bastin, Alasdair~M.J. MacLullich, Joanna~M. Wardlaw, and Ian~J. Deary.
\newblock Does white matter structure or hippocampal volume mediate associations between cortisol and cognitive ageing?
\newblock \emph{Psychoneuroendocrinology}, 62:\penalty0 129--137, December 2015.
\newblock \doi{10.1016/j.psyneuen.2015.08.005}.

\bibitem[Needleman(2004)]{Needleman2004}
Herbert Needleman.
\newblock Lead poisoning.
\newblock \emph{Annual Review of Medicine}, 55:\penalty0 209--222, 2004.
\newblock \doi{10.1146/annurev.med.55.091902.103653}.

\bibitem[Cecil et~al.(2008)Cecil, Brubaker, Adler, Dietrich, Altaye, Egelhoff, Wessel, Elangovan, Hornung, Jarvis, and Lanphear]{Cecil2008}
Kim~M Cecil, Christopher~J Brubaker, Caleb~M Adler, Kim~N Dietrich, Mekibib Altaye, John~C Egelhoff, Stephanie Wessel, Ilayaraja Elangovan, Richard Hornung, Kelly Jarvis, and Bruce~P Lanphear.
\newblock Decreased brain volume in adults with childhood lead exposure.
\newblock \emph{PLoS Medicine}, 5\penalty0 (5):\penalty0 e112, May 2008.
\newblock \doi{10.1371/journal.pmed.0050112}.

\bibitem[Lin et~al.(1998)Lin, Psaty, and Kronmal]{Lin1998Sep}
D.~Y. Lin, B.~M. Psaty, and R.~A. Kronmal.
\newblock {Assessing the sensitivity of regression results to unmeasured confounders in observational studies}.
\newblock \emph{Biometrics}, 54\penalty0 (3):\penalty0 948--963, September 1998.

\bibitem[Liu et~al.(2013)Liu, Kuramoto, and Stuart]{Liu2013Dec}
Weiwei Liu, S.~Janet Kuramoto, and Elizabeth~A. Stuart.
\newblock An introduction to sensitivity analysis for unobserved confounding in nonexperimental prevention research.
\newblock \emph{Prevention Science}, 14\penalty0 (6):\penalty0 570--580, February 2013.
\newblock \doi{10.1007/s11121-012-0339-5}.

\bibitem[Cinelli and Hazlett(2020)]{cinelli2020making}
Carlos Cinelli and Chad Hazlett.
\newblock Making sense of sensitivity: extending omitted variable bias.
\newblock \emph{Journal of the Royal Statistical Society: Series B (Statistical Methodology)}, 82\penalty0 (1):\penalty0 39--67, 2020.
\newblock \doi{10.1111/rssb.12348}.

\bibitem[Berkson(1946)]{berkson1946limitations}
Joseph Berkson.
\newblock Limitations of the application of fourfold table analysis to hospital data.
\newblock \emph{Biometrics Bulletin}, 2\penalty0 (3):\penalty0 47 -- 53, Jun 1946.
\newblock \doi{https://doi.org/10.2307/3002000}.

\bibitem[Mathur and Shpitser(2024)]{Mathur2024Jun}
Maya~B Mathur and Ilya Shpitser.
\newblock Simple graphical rules for assessing selection bias in general-population and selected-sample treatment effects.
\newblock \emph{American Journal of Epidemiology}, 194\penalty0 (1):\penalty0 267 -- 277, 2024.
\newblock \doi{10.1093/aje/kwae145}.

\bibitem[Greenland(2003)]{Greenland2003}
Sander Greenland.
\newblock Quantifying biases in causal models: classical confounding vs collider-stratification bias.
\newblock \emph{Epidemiology}, 14\penalty0 (3):\penalty0 300--306, 2003.

\bibitem[VanderWeele and Chiba(2012)]{VanderWeele2012}
Tyler~J VanderWeele and Yasutaka Chiba.
\newblock Bias formulas for sensitivity analysis for direct and indirect effects.
\newblock \emph{Epidemiology}, 23\penalty0 (1):\penalty0 42--52, 2012.

\bibitem[Kecklund and Axelsson(2016)]{Kecklund2016}
G{\"o}ran Kecklund and John Axelsson.
\newblock Health consequences of shift work and insufficient sleep.
\newblock \emph{BMJ}, 355:\penalty0 i5210, 2016.
\newblock \doi{10.1136/bmj.i5210}.

\bibitem[Torquati et~al.(2018)Torquati, Mielke, Brown, and Kolbe-Alexander]{Torquati2018}
Lucia Torquati, Gregore~I Mielke, Wendy~J Brown, and Tracy Kolbe-Alexander.
\newblock Shift work and the risk of cardiovascular disease. a systematic review and meta-analysis including dose--response relationship.
\newblock \emph{Scandinavian journal of work, environment \& health}, 44\penalty0 (3):\penalty0 229--238, 2018.

\bibitem[Biegus et~al.(2022)Biegus, Frobell, Wallin, and Ekdahl]{Biegus2022}
Karol~R. Biegus, Richard~B. Frobell, Åsa~K. Wallin, and Anne~W. Ekdahl.
\newblock The challenge of recruiting multimorbid older patients identified in a hospital database to a randomised controlled trial.
\newblock \emph{Aging Clinical and Experimental Research}, 34\penalty0 (12):\penalty0 3115–3121, October 2022.
\newblock \doi{10.1007/s40520-022-02263-0}.

\bibitem[Stern(2012)]{Stern2012}
Yaakov Stern.
\newblock Cognitive reserve in ageing and {Alzheimer's} disease.
\newblock \emph{The Lancet Neurology}, 11\penalty0 (11):\penalty0 1006--1012, 2012.
\newblock \doi{10.1016/S1474-4422(12)70191-6}.

\bibitem[Barulli and Stern(2013)]{Barulli2013}
Dana Barulli and Yaakov Stern.
\newblock Efficiency, capacity, compensation, maintenance, plasticity: emerging concepts in cognitive reserve.
\newblock \emph{Trends in cognitive sciences}, 17\penalty0 (10):\penalty0 502--509, 2013.

\bibitem[Fry et~al.(2017)Fry, Littlejohns, Sudlow, Doherty, Adamska, Sprosen, Collins, and Allen]{Fry2017}
Anna Fry, Thomas~J Littlejohns, Cathie Sudlow, Nicola Doherty, Ligia Adamska, Tim Sprosen, Rory Collins, and Naomi~E Allen.
\newblock Comparison of sociodemographic and health-related characteristics of uk biobank participants with those of the general population.
\newblock \emph{American journal of epidemiology}, 186\penalty0 (9):\penalty0 1026--1034, 2017.

\bibitem[Lee et~al.(2007)Lee, Bindman, Ford, Glozier, Moran, Stewart, and Hotopf]{Doris2021}
William Lee, Jonathan Bindman, Tamsin Ford, Nick Glozier, Paul Moran, Robert Stewart, and Matthew Hotopf.
\newblock Bias in psychiatric case-control studies.
\newblock \emph{British Journal of Psychiatry}, 190\penalty0 (3):\penalty0 204 -- 209, March 2007.
\newblock \doi{10.1192/bjp.bp.106.027250}.

\bibitem[Wachinger et~al.(2021)Wachinger, Rieckmann, and P{\"o}lsterl]{Wachinger2020}
Christian Wachinger, Anna Rieckmann, and Sebastian P{\"o}lsterl.
\newblock Detect and correct bias in multi-site neuroimaging datasets.
\newblock \emph{Medical Image Analysis}, 67:\penalty0 101879, 2021.

\bibitem[Power et~al.(2012)Power, Barnes, Snyder, Schlaggar, and Petersen]{Power2012}
Jonathan~D Power, Kelly~A Barnes, Abraham~Z Snyder, Bradley~L Schlaggar, and Steven~E Petersen.
\newblock Spurious but systematic correlations in functional connectivity mri networks arise from subject motion.
\newblock \emph{Neuroimage}, 59\penalty0 (3):\penalty0 2142--2154, 2012.

\bibitem[Alexander-Bloch et~al.(2016)Alexander-Bloch, Clasen, Stockman, Ronan, Lalonde, Giedd, and Raznahan]{Alexander2016}
Aaron Alexander-Bloch, Liv Clasen, Mark Stockman, Lisa Ronan, Francois Lalonde, Jay Giedd, and Armin Raznahan.
\newblock Subtle in-scanner motion biases automated measurement of brain anatomy from in vivo mri.
\newblock \emph{Human brain mapping}, 37\penalty0 (7):\penalty0 2385--2397, 2016.

\bibitem[Nebel et~al.(2022)Nebel, Lidstone, Wang, Benkeser, Mostofsky, and Risk]{Nebel2022Aug}
Mary~Beth Nebel, Daniel~E. Lidstone, Liwei Wang, David Benkeser, Stewart~H. Mostofsky, and Benjamin~B. Risk.
\newblock {Accounting for motion in resting-state fMRI: What part of the spectrum are we characterizing in autism spectrum disorder?}
\newblock \emph{NeuroImage}, 257:119296, August 2022.
\newblock \doi{10.1016/j.neuroimage.2022.119296}.

\bibitem[Peverill et~al.(2025)Peverill, Russell, Keding, Rich, Halvorson, King, Birn, and Herringa]{Peverill2025}
Matthew Peverill, Justin~D. Russell, Taylor~J. Keding, Hailey~M. Rich, Max~A. Halvorson, Kevin~M. King, Rasmus~M. Birn, and Ryan~J. Herringa.
\newblock Balancing data quality and bias: Investigating functional connectivity exclusions in the adolescent brain cognitive development℠ (<scp>abcd</scp> study) across quality control pathways.
\newblock \emph{Human Brain Mapping}, 46\penalty0 (1), January 2025.
\newblock \doi{10.1002/hbm.70094}.

\bibitem[Cosgrove et~al.(2022)Cosgrove, McDermott, White, Mosconi, Thompson, Paulus, Cardenas-Iniguez, and Aupperle]{Cosgrove2022Aug}
Kelly~T. Cosgrove, Timothy~J. McDermott, Evan~J. White, Matthew~W. Mosconi, Wesley~K. Thompson, Martin~P. Paulus, Carlos Cardenas-Iniguez, and Robin~L. Aupperle.
\newblock {Limits to the generalizability of resting-state functional magnetic resonance imaging studies of youth: An examination of ABCD Study{\ifmmode\circledR\else\textregistered\fi} baseline data}.
\newblock \emph{Brain imaging and behavior}, 16\penalty0 (4):\penalty0 1919 -- 1925, August 2022.
\newblock \doi{10.1007/s11682-022-00665-2}.

\bibitem[Chatfield et~al.(2005)Chatfield, Brayne, and Matthews]{Chatfield2005}
Mark~D Chatfield, Carol~E Brayne, and Fiona~E Matthews.
\newblock A systematic literature review of attrition between waves in longitudinal studies in the elderly shows a consistent pattern of dropout between differing studies.
\newblock \emph{Journal of Clinical Epidemiology}, 58\penalty0 (1):\penalty0 13--19, 2005.
\newblock \doi{10.1016/j.jclinepi.2004.05.006}.

\bibitem[Glymour et~al.(2012)Glymour, Chêne, Tzourio, and Dufouil]{Glymour2012}
M.~Maria Glymour, Geneviève Chêne, Christophe Tzourio, and Carole Dufouil.
\newblock Brain mri markers and dropout in a longitudinal study of cognitive aging.
\newblock \emph{Neurology}, 79\penalty0 (13):\penalty0 1340–1348, 2012.
\newblock \doi{10.1212/wnl.0b013e31826cd62a}.
\newblock URL \url{http://dx.doi.org/10.1212/WNL.0b013e31826cd62a}.

\bibitem[Bridgeford et~al.(2025)Bridgeford, Powell, Kiar, Noble, Chung, Panda, Lawrence, Xu, Milham, Caffo, and Vogelstein]{Bridgeford2024Feb}
Eric~W. Bridgeford, Michael Powell, Gregory Kiar, Stephanie Noble, Jaewon Chung, Sambit Panda, Ross Lawrence, Ting Xu, Michael Milham, Brian Caffo, and Joshua~T. Vogelstein.
\newblock When no answer is better than a wrong answer: A causal perspective on batch effects.
\newblock \emph{Imaging Neuroscience}, 3, 2025.
\newblock \doi{10.1162/imag_a_00458}.

\bibitem[Tripepi et~al.(2010)Tripepi, Jager, Dekker, and Zoccali]{Tripepi2010}
Giovanni Tripepi, Kitty~J Jager, Friedo~W Dekker, and Carmine Zoccali.
\newblock Selection bias and information bias in clinical research.
\newblock \emph{Nephron Clinical Practice}, 115\penalty0 (2):\penalty0 c94--c99, 2010.
\newblock \doi{10.1159/000312871}.

\bibitem[Charpentier et~al.(2021)Charpentier, Faulkner, Pool, Ly, Tollenaar, Kluen, Fransen, Yamamori, Lally, Mkrtchian, Valton, Huys, Sarigiannidis, Morrow, Krenz, Kalbe, Cremer, Zerbes, Kausche, Wanke, Giarrizzo, Pulcu, Murphy, Kaltenboeck, Browning, Paul, Cools, Roelofs, Pessoa, Harmer, Chase, Grillon, Schwabe, Roiser, Robinson, and O’Doherty]{Charpentier2021}
Caroline~J Charpentier, Paul Faulkner, Eva~R Pool, Verena Ly, Marieke~S Tollenaar, Lisa~M Kluen, Aniek Fransen, Yumeya Yamamori, Níall Lally, Anahit Mkrtchian, Vincent Valton, Quentin J~M Huys, Ioannis Sarigiannidis, Kelly~A Morrow, Valentina Krenz, Felix Kalbe, Anna Cremer, Gundula Zerbes, Franziska~M Kausche, Nadine Wanke, Alessio Giarrizzo, Erdem Pulcu, Susannah Murphy, Alexander Kaltenboeck, Michael Browning, Lynn~K Paul, Roshan Cools, Karin Roelofs, Luiz Pessoa, Catherine~J Harmer, Henry~W Chase, Christian Grillon, Lars Schwabe, Jonathan~P Roiser, Oliver~J Robinson, and John~P O’Doherty.
\newblock How representative are neuroimaging samples? large-scale evidence for trait anxiety differences between fmri and behaviour-only research participants.
\newblock \emph{Social Cognitive and Affective Neuroscience}, 16\penalty0 (10):\penalty0 1057 -- 1070, May 2021.
\newblock \doi{10.1093/scan/nsab057}.

\bibitem[Teague et~al.(2018)Teague, Youssef, Macdonald, Sciberras, Shatte, Fuller-Tyszkiewicz, Greenwood, McIntosh, Olsson, and Hutchinson]{Teague2018}
Samantha Teague, George~J. Youssef, Jacqui~A. Macdonald, Emma Sciberras, Adrian Shatte, Matthew Fuller-Tyszkiewicz, Chris Greenwood, Jennifer McIntosh, Craig~A. Olsson, and Delyse Hutchinson.
\newblock Retention strategies in longitudinal cohort studies: a systematic review and meta-analysis.
\newblock \emph{BMC Medical Research Methodology}, 18\penalty0 (1), November 2018.
\newblock \doi{10.1186/s12874-018-0586-7}.

\bibitem[Ganguli et~al.(1998)Ganguli, Lytle, Reynolds, and Dodge]{Ganguli1998}
M.~Ganguli, M.~E. Lytle, M.~D. Reynolds, and H.~H. Dodge.
\newblock Random versus volunteer selection for a community-based study.
\newblock \emph{The Journals of Gerontology Series A: Biological Sciences and Medical Sciences}, 53A\penalty0 (1):\penalty0 M39–M46, January 1998.
\newblock \doi{10.1093/gerona/53a.1.m39}.
\newblock URL \url{http://dx.doi.org/10.1093/gerona/53a.1.m39}.

\bibitem[Hern{\'a}n et~al.(2004)Hern{\'a}n, Hern{\'a}ndez-D{\'\i}az, and Robins]{hernan2004structural}
Miguel~A Hern{\'a}n, Sonia Hern{\'a}ndez-D{\'\i}az, and James~M Robins.
\newblock A structural approach to selection bias.
\newblock \emph{Epidemiology}, 15\penalty0 (5):\penalty0 615--625, 2004.

\bibitem[Rohrer(2018)]{rohrer2018thinking}
Julia~M Rohrer.
\newblock Thinking clearly about correlations and causation: Graphical causal models for observational data.
\newblock \emph{Advances in Methods and Practices in Psychological Science}, 1\penalty0 (1):\penalty0 27--42, 2018.
\newblock \doi{10.1177/2515245917745629}.

\bibitem[Karcher and Barch(2021)]{Karcher2021Jan}
Nicole~R. Karcher and Deanna~M. Barch.
\newblock {The ABCD study: understanding the development of risk for mental and physical health outcomes}.
\newblock \emph{Neuropsychopharmacology}, 46:\penalty0 131--142, January 2021.
\newblock \doi{10.1038/s41386-020-0736-6}.

\bibitem[{Anthropic}(2024)]{anthropic2024claude4}
{Anthropic}.
\newblock Claude sonnet 4.
\newblock \url{https://www.anthropic.com/claude}, 2024.
\newblock [Online. Accessed between July 1, 2025 and July 10, 2025.].

\bibitem[Bang and Robins(2005)]{bang2005doubly}
Heejung Bang and James~M Robins.
\newblock Doubly robust estimation in missing data and causal inference models.
\newblock \emph{Biometrics}, 61\penalty0 (4):\penalty0 962--973, 2005.

\bibitem[van~der Laan and Rubin(2006)]{van2006targeted}
Mark~J van~der Laan and Daniel Rubin.
\newblock Targeted maximum likelihood learning.
\newblock \emph{The International Journal of Biostatistics}, 2\penalty0 (1), 2006.

\bibitem[Ho et~al.(2011)Ho, Imai, King, and Stuart]{Ho2011Jun}
Daniel Ho, Kosuke Imai, Gary King, and Elizabeth~A. Stuart.
\newblock {MatchIt: Nonparametric Preprocessing for Parametric Causal Inference}.
\newblock \emph{Journal of Statistical Software}, 42:\penalty0 1--28, June 2011.
\newblock \doi{10.18637/jss.v042.i08}.

\bibitem[Johnson et~al.(2007)Johnson, Li, and Rabinovic]{Johnson2007Jan}
W.~Evan Johnson, Cheng Li, and Ariel Rabinovic.
\newblock {Adjusting batch effects in microarray expression data using empirical Bayes methods}.
\newblock \emph{Biostatistics}, 8\penalty0 (1):\penalty0 118--127, Jan 2007.
\newblock \doi{10.1093/biostatistics/kxj037}.

\bibitem[Friedman et~al.(2006)Friedman, Glover, and Consortium]{Friedman2006Jul}
Lee Friedman, Gary~H. Glover, and The~FBIRN Consortium.
\newblock Reducing interscanner variability of activation in a multicenter fmri study: Controlling for signal-to-fluctuation-noise-ratio (sfnr) differences.
\newblock \emph{NeuroImage}, 33\penalty0 (2):\penalty0 471–481, November 2006.
\newblock \doi{10.1016/j.neuroimage.2006.07.012}.

\bibitem[VanderWeele and Hern{\ifmmode\acute{a}\else\'{a}\fi}n(2012)]{VanderWeele2012Jun}
Tyler~J. VanderWeele and Miguel~A. Hern{\ifmmode\acute{a}\else\'{a}\fi}n.
\newblock {Results on differential and dependent measurement error of the exposure and the outcome using signed directed acyclic graphs}.
\newblock \emph{American Journal of Epidemiology}, 175\penalty0 (12):\penalty0 1303--1310, June 2012.
\newblock \doi{10.1093/aje/kwr458}.

\bibitem[Jurek et~al.(2005)Jurek, Greenland, Maldonado, and Church]{jurek2005proper}
Anne~M Jurek, Sander Greenland, George Maldonado, and Timothy~R Church.
\newblock Proper interpretation of non-differential misclassification effects: expectations vs observations.
\newblock \emph{International Journal of Epidemiology}, 34\penalty0 (3):\penalty0 680--687, 2005.

\bibitem[Lash et~al.(2009)Lash, Fox, and Fink]{lash2014applying}
Timothy~L Lash, Matthew~P Fox, and Aliza~K Fink.
\newblock \emph{Applying Quantitative Bias Analysis to Epidemiologic Data}.
\newblock Springer, New York, 1st edition, 2009.

\bibitem[Van~Dijk et~al.(2012)Van~Dijk, Sabuncu, and Buckner]{VanDijk2012Jan}
Koene R.~A. Van~Dijk, Mert~R. Sabuncu, and Randy~L. Buckner.
\newblock {The influence of head motion on intrinsic functional connectivity MRI}.
\newblock \emph{NeuroImage}, 59\penalty0 (1):\penalty0 431--438, January 2012.
\newblock \doi{10.1016/j.neuroimage.2011.07.044}.

\bibitem[Carroll et~al.(1995)Carroll, Ruppert, and Stefanski]{carroll1995measurement}
Raymond~J Carroll, David Ruppert, and Leonard~A Stefanski.
\newblock Measurement error in nonlinear models.
\newblock \emph{Chapman and Hall/CRC Monographs on Statistics and Applied Probability}, 1995.

\bibitem[le~Cessie et~al.(2012)le~Cessie, Debeij, Rosendaal, Cannegieter, and Vandenbroucke]{leCessie2012Jul}
Saskia le~Cessie, Jan Debeij, Frits~R. Rosendaal, Suzanne~C. Cannegieter, and Jan~P. Vandenbroucke.
\newblock {Quantification of Bias in Direct Effects Estimates Due to Different Types of Measurement Error in the Mediator}.
\newblock \emph{Epidemiology}, 23\penalty0 (4):\penalty0 551 -- 560, July 2012.
\newblock \doi{10.1097/EDE.0b013e318254f5de}.

\bibitem[Westfall and Yarkoni(2016)]{westfall2016statistically}
Jacob Westfall and Tal Yarkoni.
\newblock Statistically controlling for confounding constructs is harder than you think.
\newblock \emph{PLoS ONE}, 11\penalty0 (3):\penalty0 e0152719, March 2016.
\newblock \doi{10.1371/journal.pone.0152719}.

\bibitem[Feczko et~al.(2021)Feczko, Conan, Marek, Tervo-Clemmens, Cordova, Doyle, Earl, Perrone, Sturgeon, Klein, Harman, Kilamovich, Hermosillo, Miranda-Dominguez, Adebimpe, Bertolero, Cieslak, Covitz, Hendrickson, Juliano, Snider, Moore, Uriartel, Graham, Calabro, Rosenberg, Rapuano, Casey, Watts, Hagler, Thompson, Nichols, Hoffman, Luna, Garavan, Satterthwaite, Ewing, Nagel, Dosenbach, and Fair]{Feczko2021Jul}
Eric Feczko, Greg Conan, Scott Marek, Brenden Tervo-Clemmens, Michaela Cordova, Olivia Doyle, Eric Earl, Anders Perrone, Darrick Sturgeon, Rachel Klein, Gareth Harman, Dakota Kilamovich, Robert Hermosillo, Oscar Miranda-Dominguez, Azeez Adebimpe, Maxwell Bertolero, Matthew Cieslak, Sydney Covitz, Timothy Hendrickson, Anthony~C. Juliano, Kathy Snider, Lucille~A. Moore, Johnny Uriartel, Alice~M. Graham, Finn Calabro, Monica~D. Rosenberg, Kristina~M. Rapuano, B.~J. Casey, Richard Watts, Donald Hagler, Wesley~K. Thompson, Thomas~E. Nichols, Elizabeth Hoffman, Beatriz Luna, Hugh Garavan, Theodore~D. Satterthwaite, Sarah~Feldstein Ewing, Bonnie Nagel, Nico U.~F. Dosenbach, and Damien~A. Fair.
\newblock {Adolescent Brain Cognitive Development (ABCD) Community MRI Collection and Utilities}.
\newblock \emph{bioRxiv}, page 2021.07.09.451638, July 2021.
\newblock URL \url{https://doi.org/10.1101/2021.07.09.451638}.

\end{thebibliography}
